\documentclass[prl,twocolumn,superscriptaddress]{revtex4-1}
\usepackage{amsmath,amssymb,bm}
\usepackage{hyperref}
\usepackage{graphicx}
\usepackage{epstopdf}
\usepackage{latexsym}
\usepackage{subfigure}
\usepackage[usenames, dvipsnames]{color}
\usepackage[usenames, dvipsnames]{xcolor}
\usepackage{natbib}
\usepackage{braket}
\usepackage{float}
\usepackage[normalem]{ulem}
\usepackage{comment}
\usepackage{mathtools}
\usepackage{array}
\usepackage{tabu}
\usepackage{multirow}
\usepackage{chemformula}
\usepackage{svg}

\newcommand\redsout{\bgroup\markoverwith{\textcolor{red}{\rule[0.5ex]{2pt}{0.4pt}}}\ULon}
\newcommand\bluesout{\bgroup\markoverwith{\textcolor{blue}{\rule[0.5ex]{2pt}{0.4pt}}}\ULon}

\newcommand{\prlsec}[1]{{{\it #1:--}}}

\newcommand{\NDhide}[1]{{}}
\newcommand{\SPhide}[1]{{}}

\begin{document}
\title{Resummation-based Quantum Monte Carlo for quantum paramagnetic phases
}
\author{Nisheeta Desai}
\affiliation{Department of Theoretical Physics, Tata Institute of Fundamental Research,
Mumbai, MH 400005, India}
\author{Sumiran Pujari}
\email{sumiran@phy.iitb.ac.in}
\affiliation{Department of Physics, Indian Institute of
Technology Bombay, Mumbai, MH 400076, India}
\begin{abstract}
For spin rotational symmetric models with a positive-definite
high-temperature expansion of the partition function,
a stochastic sampling of the series expansion upon partial resummation becomes
logically equivalent to
sampling an uncoloured closely-packed loop-gas model in one higher dimension.
Based on this, we devise quantum Monte Carlo updates that importance-sample
loop configurations for general $SU(N)$
in fundamental and higher-symmetric representations.
The algorithmic performance systematically improves with increase in (continuous) $N$
allowing efficient simulation of quantum paramagnets.
The underlying reason for the increased efficacy is the
correspondence of quantum paramagnetic phases like valence bond solids 
to short-loop phases on the loop-gas side rather than the particular value of $N$.
This also gives a connection between Sandvik's $JQ$ model class
and classical loop-gas models in the deconfined universality class.
\end{abstract}
\maketitle

For all areas of physics including 
strongly correlated matter, efficient computational algorithms 
are now indispensable. Systematic advances in their design
have thus become the keys to progress.
Such is exemplarily the case for an important class of algorithms
based on the Monte Carlo (MC) method
that has provided unbiased insights into  
multifarious condensed matter systems, as well as lattice gauge theories
for elementary particles. 

For magnetic insulators that set our backdrop, 
quantum Monte Carlo (QMC) is now routinely used to study various
lattice quantum spin Hamiltonians \cite{Sandvik_Ch16}
which provide effective microscopic models
for the magnetically active sites in the crystal, or idealized versions
aimed at capturing the correct long-distance physics.\cite{Kaul_Melko_Sandvik_review}
For these models, an influential set of works 
~\cite{Affleck_1985,Haldane_1988,Read_Sachdev_1989a,Read_Sachdev_1989b,Read_Sachdev_1990,Read_Sachdev_1991} 
have set the agenda for charting out the landscape
of magnetic
and, importantly, quantum non-magnetic phases and associated quantum phase transitions. 
The large-$N$ perturbative approach of Refs.~\cite{Affleck_1985,Read_Sachdev_1989a,Read_Sachdev_1989b,Read_Sachdev_1990,Read_Sachdev_1991} offers insights
into $SU(2)$ magnets~\cite{Kugel_Khomskii},  
and also connects to
quantum dimer models\cite{Rokhsar_Kivelson_1988} in the $N\rightarrow \infty$ limit that serve as effective low-energy models for spin systems\cite{Read_Sachdev_1989b}.

There are two well-known flavors of 
QMC for simulating these spin models on a $d$-dimensional lattice 
at finite temperatures that are extremely efficient for small $N$. 
One is based on a path-integral representation in $d+1$ dimensions
in (imaginary) time\cite{Suzuki_1977,Beard_1996}, 
and the other based on a stochastic sampling of the 
high-temperature series expansion (SSE) of the partition function which leads
to a discrete $(d+1)$-dimensional formulation.\cite{Sandvik_1991,Sandvik_1992}
Both approaches share a
close relation --- as a simple example, there is a well-defined spin state on the lattice
($\prod_{i \in \text{lattice}} \otimes |s^z_i \rangle$) 
 at any point or slice in the additional dimension in both representations ---, 
and the ideas in one context may be ported into the 
other.\cite{Sandvik_2010_review}
There are also zero-temperature ($T=0$) projection-based QMC methods that 
stochastially project out the ground state from a trial state 
by exploiting the valence bond basis for antiferromagnetic ground states in the
singlet sector~\cite{Sandvik_2005}
including a continuous-$N$ generalization~\cite{Beach_2009}.

What has made these methods really powerful is 
the loop algorithm\cite{Evertz_1993,Ying_1993,Kawashima_1994,Beard_1996,Sandvik_1999,
Evertz_loop_review,Sandvik_2010b} 
and its extensions\cite{Prokofev_1998,Syljuasen_Sandvik_2002,Sandvik_2003,
Alet_Wessel_Troyer_2005,Heidarian_2005,Isakov_2006,Melko_2007,Biswas_2016}
which perform non-local updates 
similar to cluster updates in classical lattice 
simulations.\cite{Swendsen_Wang_1987, Wolff_1989} 
The loop algorithm is 
based on a colored loop representation of the partition function~\cite{Kaul_2015a},
and changing the color of loops leads to non-local updates. This idea can be
used in $T=0$ valence bond projector-QMC as well by reintroducing 
spin variables judiciously in the valence bond formulation
as shown by Sandvik and Evertz~\cite{Sandvik_2010b}, which also works
for Beach \textit{et al}'s continuous-$N$ generalization.~\cite{Beach_footnote}
Such loop color updates have also been exploited to study classical loop
gas models~\cite{Nahum_etal_2011,Nahum_etal_2013,Nahum_etal_2015}
whose universal properties can parallel that of spin models.
This connection goes the other way too, i.e. spin models at finite-$T$ 
may be converted to classical (uncolored) loop models in one higher 
dimension as noted in ``Suzuki-Trotterized" 
contexts.~\cite{Aizenman_1994,Kawashima_1995,Evertz_loop_review}
This is essentially a resummation over the spin variables.

In this Letter, we design finite-$T$ non-local SSE updates
based on this resummation which directly handle uncolored loops without
any reference to the underlying spin states.
This takes advantage of the basic SSE setup which incurs no 
Suzuki-Trotter errors\cite{Suzuki_1976}.
These updates lead to a systematic improvement in algorithmic
performance as $N$ increases. In relation to the classical loop-gas models
alluded to above\cite{Nahum_etal_2011,Nahum_etal_2013,Nahum_etal_2015}, 
the resultant algorithm
is well-suited for simulations of phases with predominantly short loops. 
In fact, this pure loop formulation generalizes
the essential idea of valence bond $T=0$ projector-QMC method
via the resummation-based updates
to simultaneously access both finite temperatures and any total spin sector.
More broadly speaking, uncolored short-loops in space-time are rather the natural 
objects or building blocks for
quantum non-magnetic phases like valence bond solids (VBSs). These states are of long-standing
interest both intrinsically, and  
for possible proximity to spin-liquid states. 
On the other hand, colored loops with definite spin states are the natural objects in
long-loop phases which correspond to magnetic phases that was exploited by
Sandvik-Evertz.~\cite{Sandvik_2010b} We also combine the above with the SSE methods developed by one of the authors in Ref.~\cite{Desai_2019,ND_thesis} for efficient simulation of higher-symmetric representations of $SU(N)$ that were introduced by Read and Sachdev~\cite{Read_Sachdev_1989b} to expose the myriad possible quantum non-magnetic states for higher spins.

\prlsec{Fundamental representation}
We describe the basic idea using the canonical 
$SU(2)$ spin-$\frac{1}{2}$ nearest neighbor Heisenberg Hamiltonian on a bipartite lattice. 
It is

\begin{figure}
\includegraphics[width=0.8\linewidth]{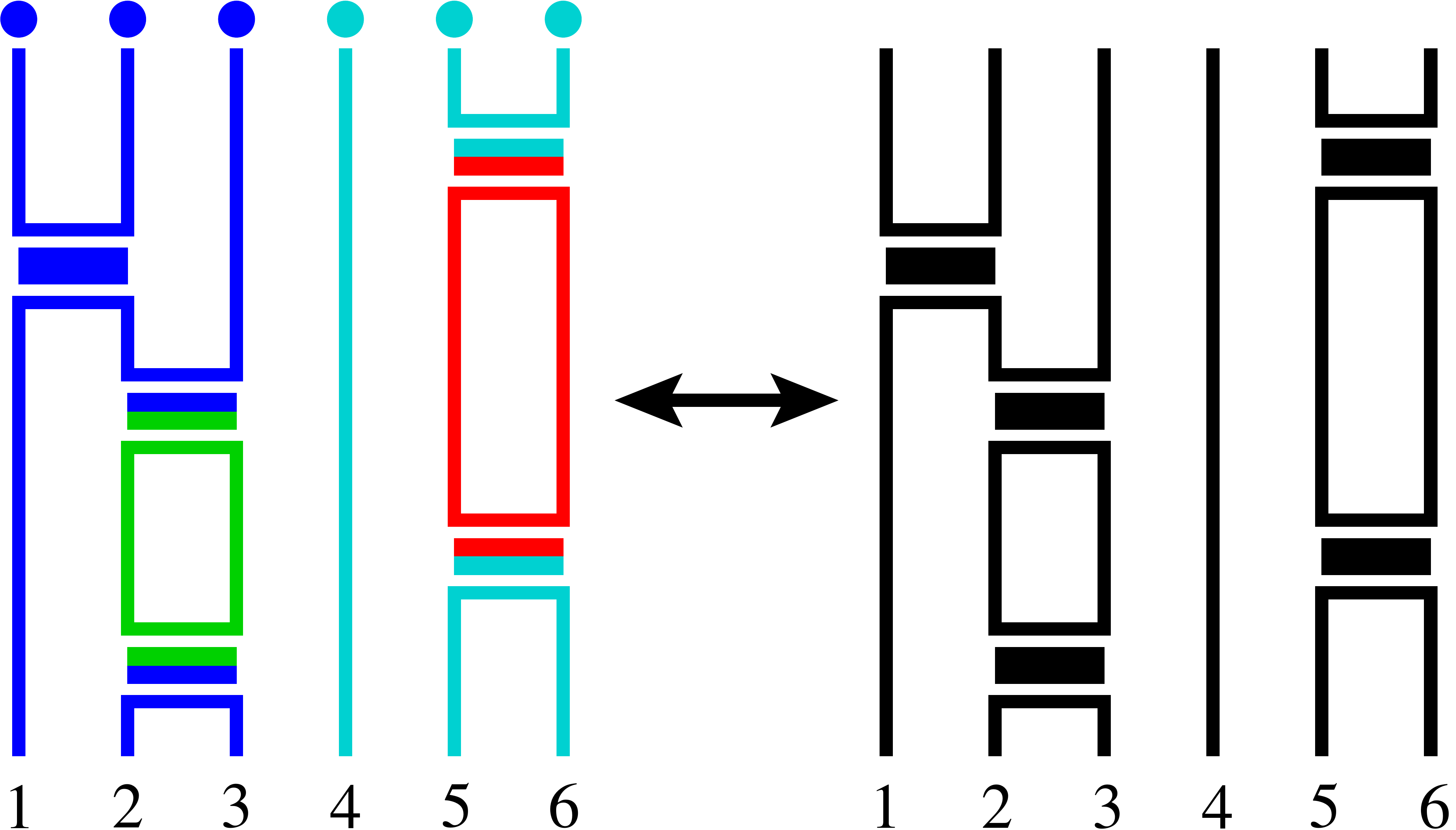}
\caption{\label{fig:diag1} Illustrative operator string configurations
for standard SSE (left) with definite spin states, and for resummed SSE (right)
characterized purely by uncolored loops.
}
\end{figure}

$H = J \sum_{\langle i,j \rangle} \mathbf{s}_i \cdot \mathbf{s}_j$
where $\mathbf{s}_i \equiv \left(s^x_i,s^y_i,s^z_i\right)$ 
are spin-$\frac{1}{2}$ operators on site $i$ at position $\mathbf{r}_i$,
and $\langle i,j \rangle$ indexes the nearest-neighbor bonds of the lattice.
After a sublattice unitary rotation, 
$H = - J \sum_{\langle i,j \rangle} H_{ij}$
with the ``singlet projector"
$H_{ij} = \frac{1}{N} \sum_{\alpha,\beta} |\alpha_i \alpha_j \rangle\langle \beta_i \beta_j| $
up to an innocuous constant.
$\alpha$,$\beta$ range from 1 to $N=2$\cite{Kaul_2011,
antifundamental_footnote}. Then, the
high-temperature series representation of the partition function
$Z(\beta) = \text{Tr} \left( e^{-\beta H} \right) = 
\sum_n \frac{(-\beta)^n}{n!} \text{Tr} \left( H^n \right)$
with $\beta \equiv \frac{1}{kT}$ becomes the (positive-definite)
operator-string representation of SSE:
\begin{equation}
\sum^{\infty}_{n=0} \frac{(-\beta)^n}{n!} \sum_{S_n} 
\sum_{\alpha} \langle \alpha | H_{\{b_1,\mu_1\}}H_{\{b_2,\mu_2\}}\hdots H_{\{b_n,\mu_n\}}| \alpha \rangle \nonumber
\end{equation}
where $S_n$ denotes a string of operator indices, and $\{b_m,\mu_m\}$ is a joint index
that tracks for the $m^{\text{th}}$ operator in the operator string its 
bond location $\langle i,j \rangle$ where $H_{ij}$ ``lives" via $b_m$,
and whether it is diagonal or off-diagonal in the usual choice of 
$s^z$ basis via $\mu_m$. 
The operator-string representation thus lives in $d+1$ dimensions. 
As remarked earlier, it can be
imagined as a configuration of closely-packed coloured loops\cite{Kaul_2015a} 
as shown
in Fig.~\ref{fig:diag1}(a).
Now, one may resum over the spin or colour values
of these closely-packed loops without breaking or changing any loop connections 
in the operator string. This 
then renders the ensemble as a configuration of closely-packed uncoloured
loops as shown in Fig.~\ref{fig:diag1}(b). This 
loop-gas representation for the high-temperature series may be written as
\begin{equation}
    Z(\beta)=\sum^{\infty}_{n=0} \frac{(-\beta)^n}{n!} N^{n_l}
    \sum_{S_n} h_{b_1} h_{b_2} \hdots h_{b_n}
    \label{eq:loopgas_rep}
\end{equation}
where bond index $b_i$ is now the only indexing required,
$h_{b_i}$ indicates the spin-symmetric matrix element 
contribution ($-\frac{J}{N}$ in our example) at $b_i$, 
and $n_l$ is the number of loops in a given configuration.

We may suggestively rewrite the above as
    $Z=   \sum_{\{ C_{\text{loops}}\}} W(C_{\text{loops}})$
where $C_{\text{loops}}$ is any allowed closely-packed uncoloured loop
gas configuration with only one underlying operator where loops
abut each other  at various
time slices (Fig.~\ref{fig:diag1}(b)), 
and $W(C_{\text{loops}}) = \frac{(\beta J/N)^n}{n!} N^{n_l}$.
The underlying operators
thus perform the role of ``transfer matrices" in the loop-gas language~\cite{Nahum_etal_2011}.
The uncolored nature of the loop-gas emerges in presence of $SU(N)$ symmetry,
which ensures that the diagonal and off-diagonal operators contribute
the same factor to the weight of the configuration.
Having done the resummation though, $N$ is now purely a parameter
and 
can be any positive, real number.
It also gets rid of the index which tracks diagonal vs. 
off-diagonal operators, which implies
a superposition of spin states at any time slice.

\prlsec{Estimators}
The simplest QMC estimator is energy, and 
it is measured in the same way here as in standard SSE. 
This is because, if we now color back the uncolored loops, 
the contribution to the energy estimator is \emph{independent} of the coloring, 
i.e. each coloring contributes the same value ($\frac{n}{\beta}$) to the energy 
estimator.~\cite{Sandvik_2010_review}
The measurement of bond operators 
also remains unchanged, e.g. $B_{\lambda}(\vec{r}) \equiv \mathbf{s}_{\mathbf{r}} \cdot 
\mathbf{s}_{\mathbf{r}+\hat{e}_\lambda}$ on the square lattice.
We can similarly measure the square lattice VBS order parameters, 
$\phi_x= \frac{1}{N_s} \sum_{\mathbf{r}}  (-1)^{r_x} 
\langle  B_x(\mathbf{r}) \rangle $ 
and $\phi_y= \frac{1}{N_s}  (-1)^{r_y} 
\langle B_y(\mathbf{r}) \rangle $. 
Measuring the correlations of the bond operator 
$\tilde{C}_{\phi^2_\lambda}(\mathbf{r})= \langle B_\lambda({\mathbf{0}})B_\lambda({\mathbf{r}})\rangle$ 
is also straightforward. 
The estimator of the spin stiffness which tracks magnetic ordering
is changed due to the resummation. 
In standard SSE, the stiffness is related to the winding of colored loops 
according to the following relation
 $\rho=\frac{\langle \mathcal{W}^2_c \rangle}{\beta}$,
where $\langle \mathcal{W}^2_c \rangle$ is the winding fluctuations of colored 
loops.~\cite{Sandvik_2010_review}
It is related to the winding fluctuations of uncolored loops as
 $\langle \mathcal{W}^2_u \rangle = \frac{N^2}{(N-1)}\langle  \mathcal{W}^2_c \rangle$
which gives an ``improved" estimator for the stiffness:
\begin{equation}
 \rho=\frac{(N-1)}{N^2} \frac{\langle \mathcal{W}^2_u \rangle}{\beta}.
\end{equation}
The derivation of the winding
fluctation relation is given in Ref.\cite{supp}. This estimator for stiffness
can be used in standard SSE as well.

\begin{figure}[t]
\includegraphics[width=\linewidth,clip=true,trim=10mm 0mm 0mm 0mm]{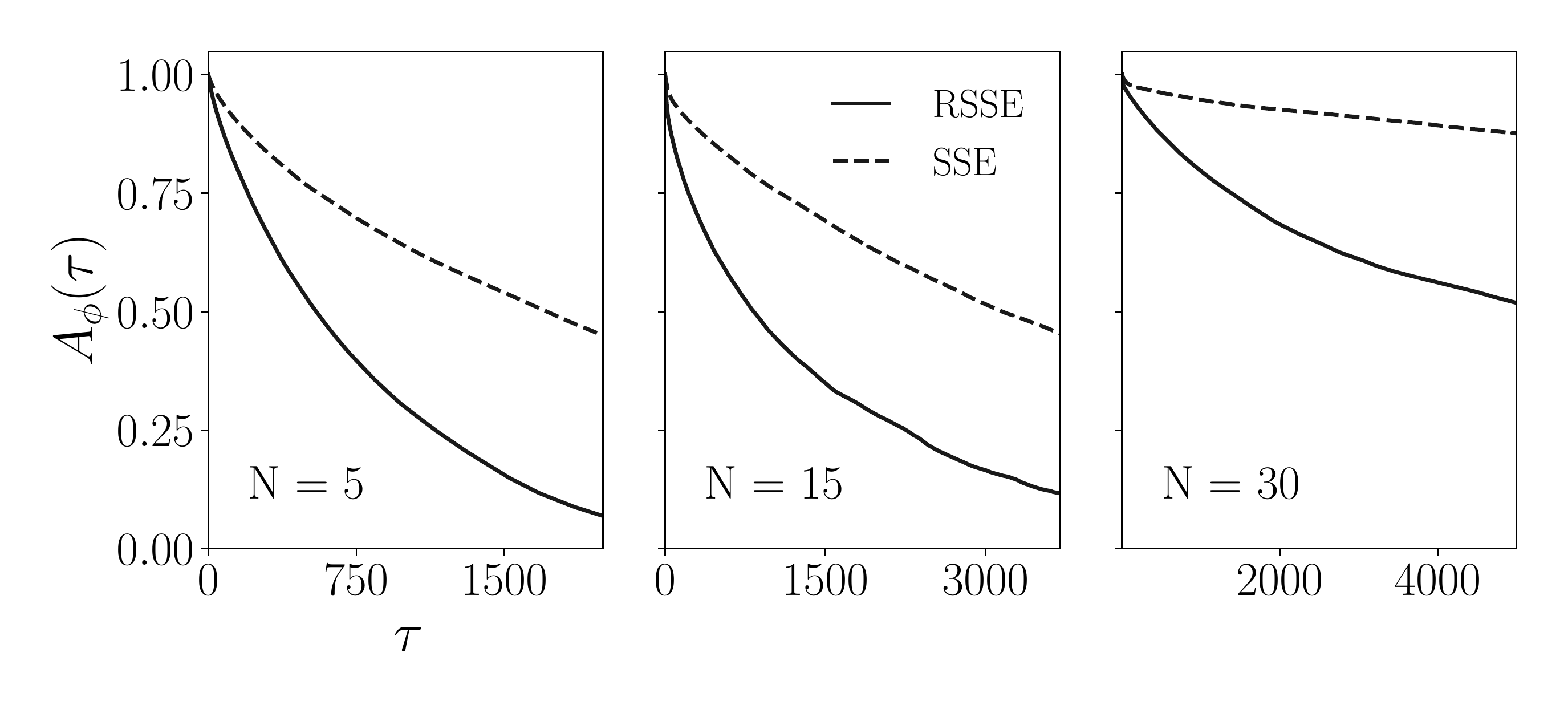}
 \includegraphics[width=\linewidth,clip=true,trim=10mm 4mm 10mm 5.2mm]{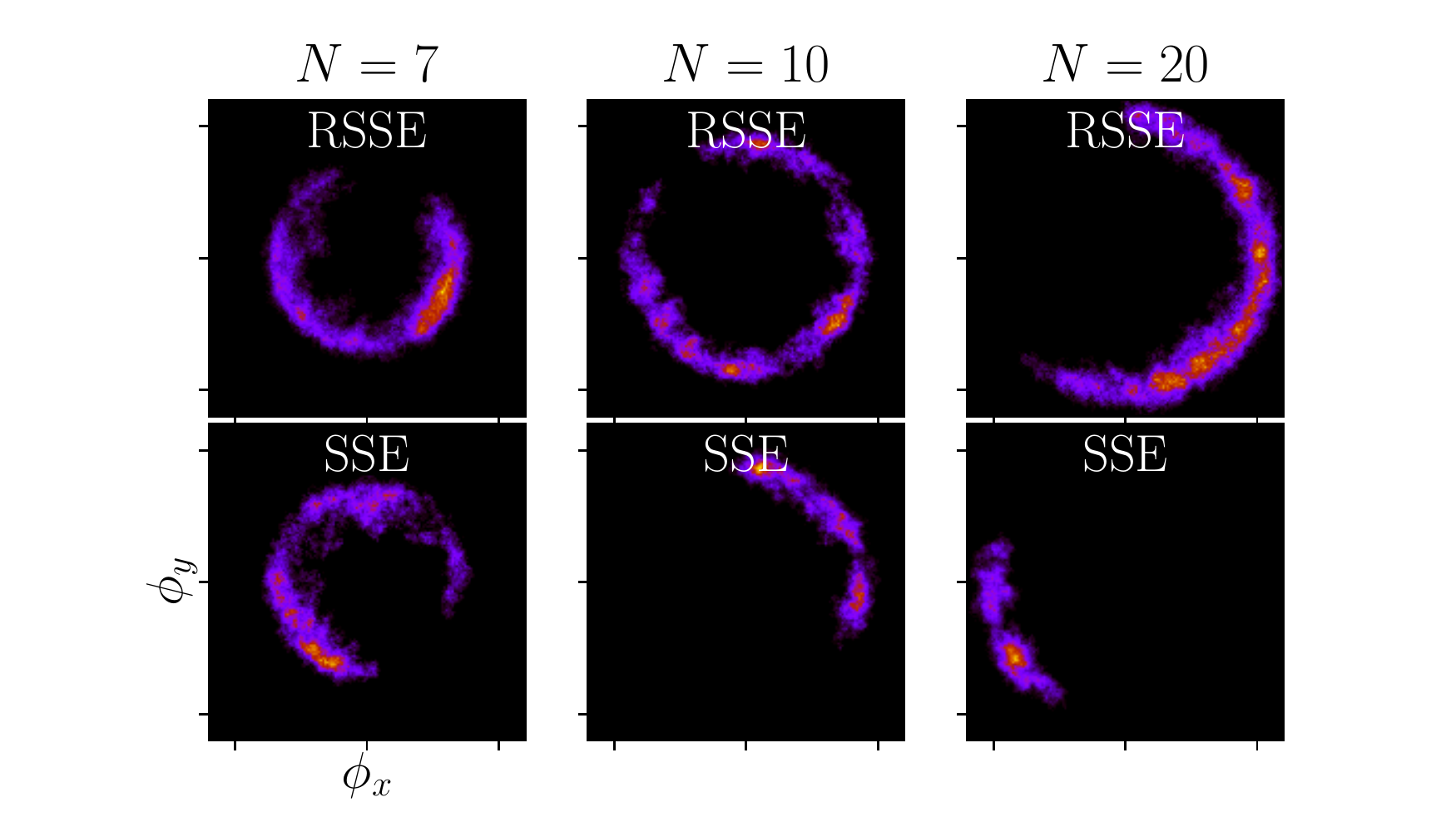}
 \caption{\label{fig:histcomparison} (top)
Comparison of autocorrelation of the VBS order parameter ($\phi=\sqrt{\phi_x^2+\phi_y^2}$) on a $16 \times 16$ lattice for $N=5, 15, 30$ at $\beta=128,64,64$ respectively measured after each Monte Carlo step using the resummation algorithm with that of the standard SSE. Autocorrelations fall off faster in the resummed SSE (RSSE) algorithm systematically as $N$ increases.
These values of $N$ correspond to the short-loop or VBS phase\cite{kawashima_harada_troyer_prl2003}
showing the efficacy of resummation-based updates 
for short-loop phases in general.
(bottom) Joint histograms of the VBS order parameters, $\phi_x$ and $\phi_y$, for $N=7,10,20$, 
on a $16 \times 16$ lattice at $\beta=128, 128, 64$ respectively. $\phi_x$ and $\phi_y$ have been measured at 
every Monte Carlo step for $10^6$ steps. 
For $N=7$, the performance of both algorithms 
are comparable. However, 
as we go deeper into the VBS phase for higher values of $N$, 
we clearly see that resummed SSE (RSSE) spans the angular 
space of the histogram better, 
thus proving to be the more ergodic algorithm in this regime.}
\end{figure}

\prlsec{Implementation and Application}
Based on Eq.~\ref{eq:loopgas_rep}, we 
implement a Monte Carlo algorithm to 
directly sample the uncoloured loop ensemble.
Each MC update consists of proposing to insert a spin-symmetric operator at an “identity” location or
to remove an already existing “non-identity” spin-symmetric operator at various space-time locations. 
The proposals get accepted with Metropolis probabilities which are
governed by the change $n \rightarrow n \pm 1$ (for Heisenberg model)
and the change in the number of loops $\delta (n_l)$.~\cite{supp}
For $H_{ij}$,
$\delta (n_l)$
takes only the values $\pm 1$ purely
due to considerations of loop topology.

One may anticipate improved performance as $N$ increases compared
to standard SSE: in any (typical) instance of the 
standard SSE operator string configuration, the off-diagonal
operator contributions start to dominate those
of diagonal operators as $N$ increases. This is simply 
due to there being $N(N-1)$ off-diagonal operators vs. $N$ diagonal
operators in $H_{ij}$.
This makes the ``diagonal" update of standard SSE
-- that changes $n$ by inserting or removing diagonal operators between
two (identical) definite spin states -- 
less efficient in updating
the operator string (the ``off-diagonal" update of standard SSE changes only
the spin or color value of the loop as remarked earlier, and does not change $n$). 
This is a non-issue in our algorithm; operators
can be potentially inserted or removed at any space-time location.
This improved performance is indeed seen as discussed in 
Fig.~\ref{fig:histcomparison}.
From a loop-gas perspective on the other hand, the resummed SSE algorithm is apt 
for simulating any phase with predominanty short loops 
where the computation of $\delta (n_l)$ becomes quite efficient. 
This is the underlying reason to prefer resummed SSE 
for short-loop phases (regardless of $N$)
like VBS (Fig.~\ref{fig:histcomparison}) as mentioned earlier. 
This should apply for other non-magnetic phases like the plaquette VBS or 
the Haldane-nematic phase.~\cite{Affleck_1988,Read_Sachdev_1990} 
These updates can also supplement the
standard SSE diagonal and loop updates in long-loop
phases or near transitions if needed for performance.

\begin{figure}[t]
\includegraphics[width=0.95\linewidth,clip=true,trim=0 5mm 0 5mm]{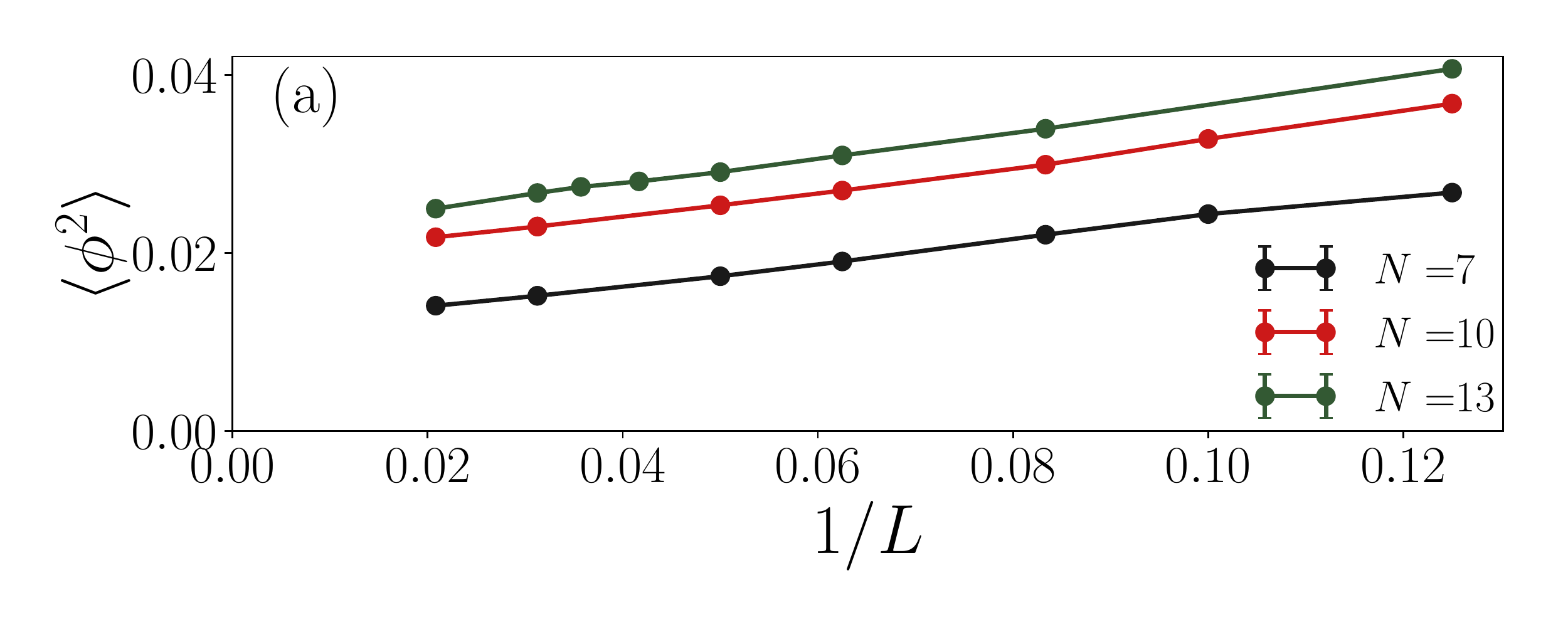}
\includegraphics[width=0.90\linewidth,clip=true,trim=10mm 30mm 10mm 30mm]{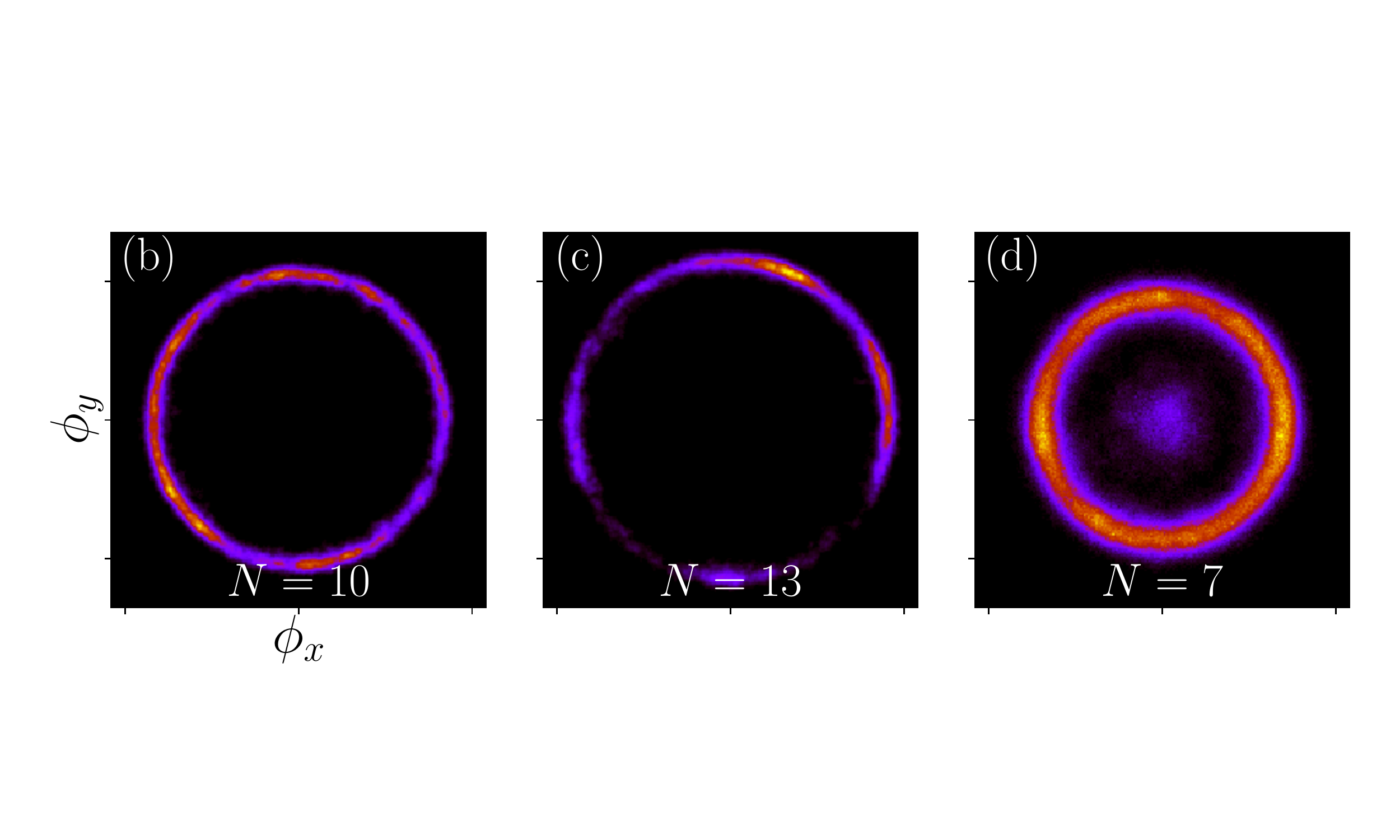}
\caption{ \label{fig:applications} (a) Finite size scaling of the VBS order parameter 
for the $SU(N)$ square lattice antiferromagnet in the
fundamental representation for $N=7,10,13$ 
at $\beta=128$ representative of zero-temperature limit~\cite{supp,error_footnote}. 
It clearly extrapolates to finite values showing VBS order in this regime 
as is expected\cite{kawashima_harada_troyer_prl2003}. (bottom)  The joint histogram of $\phi_x$ and $\phi_y$ measured at every 10$^{\text{th}}$ Monte Carlo step with a total of $10^7$ steps.
(b)-(c) At low enough temperatures, the histograms show $U(1)$ symmetry for $N=10,13$ (at $\beta=128,256$ respectively) 
on a $32 \times 32$ lattice. 
(d) For $N=7$, at $\beta=64$ on a $16 \times 16$ lattice, the distribution peaks at an U(1) symmetric ring and at $\phi_x=\phi_y=0$ suggesting a first-order transition. 
}
\end{figure}

We now apply the resummed SSE algorithm to the square lattice
antiferromagnet, $H = -\frac{J}{N} \sum_{\langle ij \rangle} \sum^{N}_{\alpha,\beta=1} |\alpha_i \alpha_j \rangle\langle \beta_i \beta_j|$, for several $N$. In the fundamental representation,
it maps to the non-interacting quantum dimer model (QDM) with one dimer
per vertex as $N \rightarrow \infty$\cite{Read_Sachdev_1989b}.
Only very recently, an efficient
algorithm based on SSE has been developed for finite-$T$ simulations
directly in the constrained Hilbert space of the QDM.\cite{sweeping_cluster_2019}
We can also access the large-$N$ regime in our simulations efficiently. 
In Fig.~\ref{fig:applications}(a), we see VBS order at low
enough temperatures in this regime. 
However, we see $U(1)$-symmetric VBS order histograms as shown in 
Fig.~\ref{fig:applications}(b,c) in contrast to the ``mixed" phase histograms 
of Ref.\cite{Yan_etal_2021} for similar system sizes.
The approach to the constrained Hilbert space of QDM in the loop
gas representation can also be quantified as shown in the final section of Ref.~\onlinecite{supp}.
We find non-negligible deviations from QDM Hilbert space to put $SU(N)$ magnets
away from the perturbative neighborhood of QDM  
even for quite large $N$. We ascribe this to
the contrast between our results and Ref.\cite{Yan_etal_2021} --
a relevant detail on the connection between $SU(N)$ magnets and QDM.
Fig.~\ref{fig:applications}(d) shows how the algorithm perfoms
near the thermal transition out of the ordered phase. 

\begin{figure}[t]
\includegraphics[width=0.9\linewidth, angle=0]{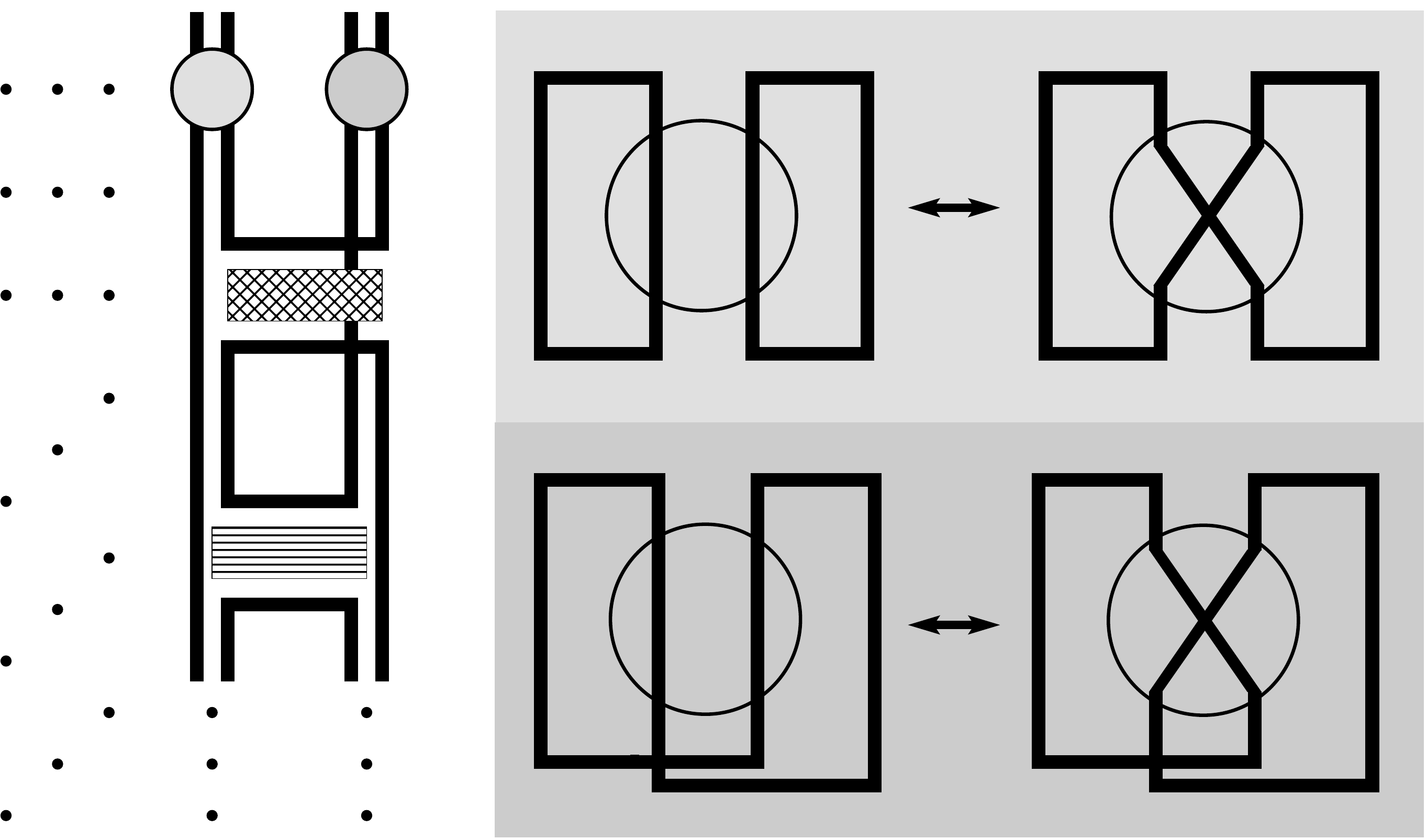}
\caption{ \label{fig:proj_diag}
Schematic to illustrate the resummation over the projection operators.
On the left side is shown a small section of the uncolored loop QMC
configuration with two split-spins per site. On the right is shown the
re-wiring MC update at the projection time slice. The shade of gray
corresponds to the projection operators shown as circles
in the QMC configuration on the left, assuming a loop geometry 
without any other operators on the shown bond elsewhere in time.
}
\end{figure}

\prlsec{Higher symmetric representations}
We now write down resummation-based updates for higher representations by making use
of the ``split-spin" language\cite{Kawashima_1995,Todo_2001,
Kawashima_2004,Desai_2019} 
which splits $\mathbf{S}_i$ as
$\mathcal{P}_i \left( \sum^{2S}_{a=1} \mathbf{s}_{i,a} \right) \mathcal{P}_i$
where $\mathcal{P}_i$ are appropriate projection operators to stay in the correct
Hilbert space.
We take the spin-$1$ Heisenberg model for $SU(2)$ as our example,
which automatically extends to the $SU(N)$ case with two symmetric ``flavours".
The spin-$1$ Heisenberg model 
$H = J \sum_{\langle i,j \rangle} \left(\mathbf{S}_i \cdot \mathbf{S}_j \right)$ 
in the split-spin language is written as 
$H = \mathcal{P} \tilde{H} \mathcal{P}$ with
$\tilde{H} = J
\sum_{\langle i,j \rangle} \sum_{a,b} \mathbf{s}_{i,a} \cdot \mathbf{s}_{j,b}$
and $a,b$ is now a split-spin index running over the number of
symmetric flavors. $\mathcal{P}$ is a projection operator that
projects onto fully symmetric subspace over the split-spins, i.e.
$\mathcal{P} = \prod_i \mathcal{P}_i$ and
$\mathcal{P}_i \equiv 
|\uparrow \uparrow \rangle \langle \uparrow \uparrow|
+ |\downarrow \downarrow \rangle \langle \downarrow \downarrow|
+ \left( \frac{|\uparrow \downarrow \rangle + |\downarrow \uparrow \rangle}
{\sqrt{2}} \right)
\left( \frac{\langle \uparrow \downarrow | + \langle \downarrow \uparrow |}
{\sqrt{2}} \right)$ for the two split-spins on the $i^\text{th}$ site.
With this in hand, the standard-SSE operator-string
representation follows from $Z(\beta) = \text{Tr}_{\mathbf{S}} 
\left( e^{-\beta H} \right) = 
\text{Tr}_{\mathbf{s}} \left( e^{-\beta \tilde{H}} \mathcal{P} \right)$.
To ensure symmetrization, it is enough that the projection operator acts
at one particular time slice.
\cite{Todo_2001} We may now 
resum as before to get a configuration in terms of uncoloured loops with, in this case, two “parallel” loops running at each space-time point
as sketched on the left side of Fig.\ref{fig:proj_diag}. 
The resummation over colours proceeds exactly the same as before 
to give the $N^{n_l}$ re-weighting
factor. 

\begin{figure}[t]
 \includegraphics[width=0.99\linewidth,clip=true,trim=9mm 10mm 5mm 0]{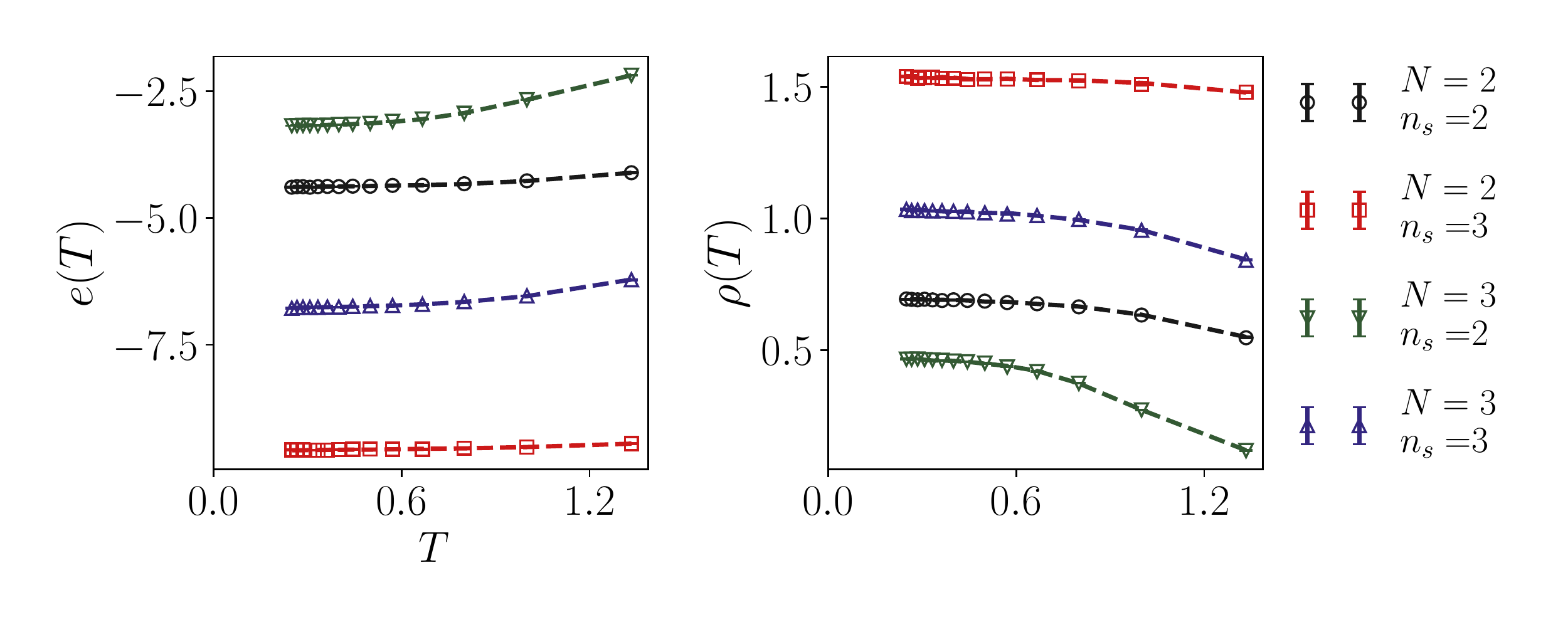}
 \caption{ \label{fig:minispin_check}
Comparison of finite temperature energy per unit site, $e(T)$, and spin stiffness, $\rho(T)$, measured in our algorithm with standard SSE (dashed lines) on a $4 \times 4$ lattice for several $N$ and $n_s$ (number of split-spins or symmetric flavors)~\cite{error_footnote}. 
}
\end{figure}

To resum over the 
projection operator, one must ensure that
the action of the projection operator $\mathcal{P}_i$ is faithfully captured
on all sites. 
In standard SSE, one implements the projection by
the use of a directed loop update\cite{Syljuasen_Sandvik_2002,Alet_Wessel_Troyer_2005} 
at the projection time slice.\cite{footnote_ND} 
Resumming over this now amounts to a ``re-wiring" of the two uncolored
loops at the two split-spin sites at this time slice as shown on the right side
of Fig.~\ref{fig:proj_diag}.
So, the implementation essentially mimics that of the earlier section
with an extra MC update for the projection time slice where one proposes
to reconfigure the split-spin connections as in Fig.~\ref{fig:proj_diag}
with the acceptance probability again governed by the $\delta(n_l)$
due to this proposal.\cite{supp}
Extending this to yet higher-symmetric representations
follows along very
similar lines with more split-spins per site. 
In Fig.~\ref{fig:minispin_check}, 
we demonstrate the implementation of these extensions 
by benchmarking energy and spin stiffness with standard SSE. Simulations of $SU(N)$ models in these higher-symmetric representations have previously been carried out using colored loop updates~\cite{kawashima_tanabe_prl2007,okubo_etal_prb2015}. Our algorithm provides an alternative to this that can allow more efficient access to the quantum non-magnetic states of $SU(N)$ antiferromagnets~\cite{Read_Sachdev_1989b,Read_Sachdev_1990}
thereby opening up further studies on (quantum)
phase transitions with $N$ and $T$.\cite{proj_footnote}

\prlsec{Discussion}
One may finally ask if there is
a physical meaning to the uncolored loops, or are they just algorithmic constructs.
Any such interpretation, apart from providing intuition, can help formulate
other useful QMC estimators~\cite{Beach_2006}
based on our generalization of the valence bond $T=0$ projector-QMC
method to finite-$T$ without the singlet sector restriction.
We can indeed interpret them as follows: at a coloured level in standard SSE,
an operator 
$|\alpha_i \alpha_j\rangle\langle \beta_i \beta_j|$ 
destroys $\beta_i$,$\beta_j$ in a state 
(``below" the operator in Fig.~\ref{fig:diag1}) and
creates $\alpha_i$,$\alpha_j$ in the resultant state 
(``above" the operator). 
Thus, $|\alpha_i \alpha_j\rangle\langle \beta_i \beta_j| \equiv 
b^\dagger_{i,\alpha} b^\dagger_{j,\alpha} b_{i,\beta} b_{j,\beta}$ where
$b$, $b^\dagger$ stand for the destruction and creation operations respectively. If 
$\chi_{ij,\alpha} \equiv b_{i,\alpha} b_{j,\alpha}$, then an $\alpha$-colored loop
is of the form 
$\chi^\dagger_{li,\alpha} \ldots \chi^\dagger_{jk,\alpha}  \chi_{ij,\alpha}$.
Therefore, upon resumming, an uncolored loop has the following $SU(N)$-symmetric
operator content:
$\tilde{\chi}^\dagger_{li} \ldots \tilde{\chi}^\dagger_{jk}  \tilde{\chi}_{ij}$ with 
$\tilde{\chi}_{ij} = \sum_\alpha \chi_{ij,\alpha}$. 
A similar interpretation
in the ``reverse" direction from loops to magnetic degrees of freedom was laid
down by Nahum \textit{et al} in Refs.\cite{Nahum_etal_2011,Nahum_etal_2013,Nahum_PRB_IIIA}, but
the Hamiltonian does not take a simple form for $3d$ loop-gases\cite{Nahum_PRB_IIID}.
Resummation instead gives a recipe to go from local Hamiltonians
to loop-gases including for higher-symmetric representations.

The above interpretation gives a connection between the $2d$
$JQ$ models\cite{Sandvik_2007,Kaul_Melko_Sandvik_review}
and the $3d$ loop-gas model studied by Nahum \textit{et al} in Ref.\cite{Nahum_etal_2015} 
both of which have been argued
to exhibit deconfined criticality\cite{Senthil_etal_2004a,Senthil_etal_2004b}. 
The $Q$ term -- a tensor 
product of several singlet projectors over independent bonds 
$\prod^p_{i=1} H_{b_i}$ -- gives
an additional rule for how loops may abut each other on a given time slice.
A $Q$-operator in the uncoloured loop representation will lead to $p$ loop abutments at a given time 
slice of the layered extension of the underlying lattice,
just as each $H_b$ led to one loop abutment as in Fig.~\ref{fig:diag1}.
The $J$,$Q$ terms in the loop-gas language thus define appropriate
transfer matrices. 
This throws a ``forward" perspective on Ref.\cite{Nahum_etal_2015,Nahum_PRX_lattice}, 
in that by resumming, any spin-symmetric Hamiltonian with a positive-definite
high-temperature series expansion exhibiting 
deconfined criticality implies the same between a long-loop and a short-loop phase
in a logically equivalent loop-gas.

\prlsec{Acknowledgements}
We thank Fabien Alet, Kedar Damle and Ribhu Kaul for discussions.
We especially thank Kedar Damle for a critical reading and feedback on the manuscript.
The numerical results were obtained using 
the computational facilities (Chandra cluster) of the 
Department of Physics, IIT Bombay.
N.D. was partially supported by NSF grant DMR-1611161,
and partially by a National Postdoctoral Fellowship of
SERB, DST, Govt. of India (PDF/2020/001658) 
at the Dept of Theoretical Physics, TIFR.
S.P. was supported by IRCC, IIT Bombay (17IRCCSG011) and SERB, DST, India
(SRG/2019/001419).

\bibliographystyle{apsrev}
\bibliography{resum_refs}

\begin{thebibliography}{64}
\expandafter\ifx\csname natexlab\endcsname\relax\def\natexlab#1{#1}\fi
\expandafter\ifx\csname bibnamefont\endcsname\relax
  \def\bibnamefont#1{#1}\fi
\expandafter\ifx\csname bibfnamefont\endcsname\relax
  \def\bibfnamefont#1{#1}\fi
\expandafter\ifx\csname citenamefont\endcsname\relax
  \def\citenamefont#1{#1}\fi
\expandafter\ifx\csname url\endcsname\relax
  \def\url#1{\texttt{#1}}\fi
\expandafter\ifx\csname urlprefix\endcsname\relax\def\urlprefix{URL }\fi
\providecommand{\bibinfo}[2]{#2}
\providecommand{\eprint}[2][]{\url{#2}}

\bibitem[{\citenamefont{{Sandvik}}(2019)}]{Sandvik_Ch16}
\bibinfo{author}{\bibfnamefont{A.~W.} \bibnamefont{{Sandvik}}},
  \bibinfo{organization}{Autumn School on Correlated Electrons, Jülich
  (Germany), 16 Sep 2019 - 20 Sep 2019} (\bibinfo{publisher}{Forschungszentrum
  Jülich GmbH Zentralbibliothek, Verlag}, \bibinfo{address}{Jülich},
  \bibinfo{year}{2019}), vol.~\bibinfo{volume}{9} of
  \emph{\bibinfo{series}{Schriften des Forschungszentrums Jülich. Modeling and
  Simulation}}, pp. \bibinfo{pages}{16.1--16.32}, ISBN
  \bibinfo{isbn}{978-3-95806-400-3},
  \urlprefix\url{https://juser.fz-juelich.de/record/864818}.

\bibitem[{\citenamefont{Kaul et~al.}(2013)\citenamefont{Kaul, Melko, and
  Sandvik}}]{Kaul_Melko_Sandvik_review}
\bibinfo{author}{\bibfnamefont{R.~K.} \bibnamefont{Kaul}},
  \bibinfo{author}{\bibfnamefont{R.~G.} \bibnamefont{Melko}}, \bibnamefont{and}
  \bibinfo{author}{\bibfnamefont{A.~W.} \bibnamefont{Sandvik}},
  \bibinfo{journal}{Annual Review of Condensed Matter Physics}
  \textbf{\bibinfo{volume}{4}}, \bibinfo{pages}{179} (\bibinfo{year}{2013}),
  \eprint{https://doi.org/10.1146/annurev-conmatphys-030212-184215},
  \urlprefix\url{https://doi.org/10.1146/annurev-conmatphys-030212-184215}.

\bibitem[{\citenamefont{Affleck}(1985)}]{Affleck_1985}
\bibinfo{author}{\bibfnamefont{I.}~\bibnamefont{Affleck}},
  \bibinfo{journal}{Phys. Rev. Lett.} \textbf{\bibinfo{volume}{54}},
  \bibinfo{pages}{966} (\bibinfo{year}{1985}),
  \urlprefix\url{https://link.aps.org/doi/10.1103/PhysRevLett.54.966}.

\bibitem[{\citenamefont{Haldane}(1988)}]{Haldane_1988}
\bibinfo{author}{\bibfnamefont{F.~D.~M.} \bibnamefont{Haldane}},
  \bibinfo{journal}{Phys. Rev. Lett.} \textbf{\bibinfo{volume}{61}},
  \bibinfo{pages}{1029} (\bibinfo{year}{1988}),
  \urlprefix\url{https://link.aps.org/doi/10.1103/PhysRevLett.61.1029}.

\bibitem[{\citenamefont{Read and
  Sachdev}(1989{\natexlab{a}})}]{Read_Sachdev_1989a}
\bibinfo{author}{\bibfnamefont{N.}~\bibnamefont{Read}} \bibnamefont{and}
  \bibinfo{author}{\bibfnamefont{S.}~\bibnamefont{Sachdev}},
  \bibinfo{journal}{Phys. Rev. Lett.} \textbf{\bibinfo{volume}{62}},
  \bibinfo{pages}{1694} (\bibinfo{year}{1989}{\natexlab{a}}),
  \urlprefix\url{https://link.aps.org/doi/10.1103/PhysRevLett.62.1694}.

\bibitem[{\citenamefont{Read and
  Sachdev}(1989{\natexlab{b}})}]{Read_Sachdev_1989b}
\bibinfo{author}{\bibfnamefont{N.}~\bibnamefont{Read}} \bibnamefont{and}
  \bibinfo{author}{\bibfnamefont{S.}~\bibnamefont{Sachdev}},
  \bibinfo{journal}{Nuclear Physics B} \textbf{\bibinfo{volume}{316}},
  \bibinfo{pages}{609} (\bibinfo{year}{1989}{\natexlab{b}}), ISSN
  \bibinfo{issn}{0550-3213},
  \urlprefix\url{https://www.sciencedirect.com/science/article/pii/0550321389900618}.

\bibitem[{\citenamefont{Read and Sachdev}(1990)}]{Read_Sachdev_1990}
\bibinfo{author}{\bibfnamefont{N.}~\bibnamefont{Read}} \bibnamefont{and}
  \bibinfo{author}{\bibfnamefont{S.}~\bibnamefont{Sachdev}},
  \bibinfo{journal}{Phys. Rev. B} \textbf{\bibinfo{volume}{42}},
  \bibinfo{pages}{4568} (\bibinfo{year}{1990}),
  \urlprefix\url{https://link.aps.org/doi/10.1103/PhysRevB.42.4568}.

\bibitem[{\citenamefont{Read and Sachdev}(1991)}]{Read_Sachdev_1991}
\bibinfo{author}{\bibfnamefont{N.}~\bibnamefont{Read}} \bibnamefont{and}
  \bibinfo{author}{\bibfnamefont{S.}~\bibnamefont{Sachdev}},
  \bibinfo{journal}{Phys. Rev. Lett.} \textbf{\bibinfo{volume}{66}},
  \bibinfo{pages}{1773} (\bibinfo{year}{1991}),
  \urlprefix\url{https://link.aps.org/doi/10.1103/PhysRevLett.66.1773}.

\bibitem[{Kug()}]{Kugel_Khomskii}
\bibinfo{note}{Also, $SU(N)$ or, more generally, $SU(2) \bigotimes SU(M)$ arise
  physically for Kugel-Khomskii models with both spin and orbital degrees of
  freedom, and other contexts with additional internal quantum numbers. The
  (algorithmic) ideas of this paper extend to these models as well.}

\bibitem[{\citenamefont{Rokhsar and Kivelson}(1988)}]{Rokhsar_Kivelson_1988}
\bibinfo{author}{\bibfnamefont{D.~S.} \bibnamefont{Rokhsar}} \bibnamefont{and}
  \bibinfo{author}{\bibfnamefont{S.~A.} \bibnamefont{Kivelson}},
  \bibinfo{journal}{Phys. Rev. Lett.} \textbf{\bibinfo{volume}{61}},
  \bibinfo{pages}{2376} (\bibinfo{year}{1988}),
  \urlprefix\url{https://link.aps.org/doi/10.1103/PhysRevLett.61.2376}.

\bibitem[{\citenamefont{Suzuki et~al.}(1977)\citenamefont{Suzuki, Miyashita,
  and Kuroda}}]{Suzuki_1977}
\bibinfo{author}{\bibfnamefont{M.}~\bibnamefont{Suzuki}},
  \bibinfo{author}{\bibfnamefont{S.}~\bibnamefont{Miyashita}},
  \bibnamefont{and} \bibinfo{author}{\bibfnamefont{A.}~\bibnamefont{Kuroda}},
  \bibinfo{journal}{Progress of Theoretical Physics}
  \textbf{\bibinfo{volume}{58}}, \bibinfo{pages}{1377} (\bibinfo{year}{1977}),
  ISSN \bibinfo{issn}{0033-068X},
  \eprint{https://academic.oup.com/ptp/article-pdf/58/5/1377/5389651/58-5-1377.pdf},
  \urlprefix\url{https://doi.org/10.1143/PTP.58.1377}.

\bibitem[{\citenamefont{Beard and Wiese}(1996)}]{Beard_1996}
\bibinfo{author}{\bibfnamefont{B.~B.} \bibnamefont{Beard}} \bibnamefont{and}
  \bibinfo{author}{\bibfnamefont{U.-J.} \bibnamefont{Wiese}},
  \bibinfo{journal}{Phys. Rev. Lett.} \textbf{\bibinfo{volume}{77}},
  \bibinfo{pages}{5130} (\bibinfo{year}{1996}),
  \urlprefix\url{https://link.aps.org/doi/10.1103/PhysRevLett.77.5130}.

\bibitem[{\citenamefont{Sandvik and Kurkij\"arvi}(1991)}]{Sandvik_1991}
\bibinfo{author}{\bibfnamefont{A.~W.} \bibnamefont{Sandvik}} \bibnamefont{and}
  \bibinfo{author}{\bibfnamefont{J.}~\bibnamefont{Kurkij\"arvi}},
  \bibinfo{journal}{Phys. Rev. B} \textbf{\bibinfo{volume}{43}},
  \bibinfo{pages}{5950} (\bibinfo{year}{1991}),
  \urlprefix\url{https://link.aps.org/doi/10.1103/PhysRevB.43.5950}.

\bibitem[{\citenamefont{Sandvik}(1992)}]{Sandvik_1992}
\bibinfo{author}{\bibfnamefont{A.~W.} \bibnamefont{Sandvik}},
  \bibinfo{journal}{Journal of Physics A: Mathematical and General}
  \textbf{\bibinfo{volume}{25}}, \bibinfo{pages}{3667} (\bibinfo{year}{1992}),
  \urlprefix\url{https://doi.org/10.1088/0305-4470/25/13/017}.

\bibitem[{\citenamefont{{Sandvik}}(2010)}]{Sandvik_2010_review}
\bibinfo{author}{\bibfnamefont{A.~W.} \bibnamefont{{Sandvik}}}, in
  \emph{\bibinfo{booktitle}{Lectures on the Physics of Strongly Correlated
  Systems Xiv: Fourteenth Training Course in the Physics of Strongly Correlated
  Systems}}, edited by
  \bibinfo{editor}{\bibfnamefont{A.}~\bibnamefont{{Avella}}} \bibnamefont{and}
  \bibinfo{editor}{\bibfnamefont{F.}~\bibnamefont{{Mancini}}}
  (\bibinfo{year}{2010}), vol. \bibinfo{volume}{1297} of
  \emph{\bibinfo{series}{American Institute of Physics Conference Series}}, pp.
  \bibinfo{pages}{135--338}, \eprint{1101.3281}.

\bibitem[{\citenamefont{Sandvik}(2005)}]{Sandvik_2005}
\bibinfo{author}{\bibfnamefont{A.~W.} \bibnamefont{Sandvik}},
  \bibinfo{journal}{Phys. Rev. Lett.} \textbf{\bibinfo{volume}{95}},
  \bibinfo{pages}{207203} (\bibinfo{year}{2005}),
  \urlprefix\url{https://link.aps.org/doi/10.1103/PhysRevLett.95.207203}.

\bibitem[{\citenamefont{Beach et~al.}(2009)\citenamefont{Beach, Alet, Mambrini,
  and Capponi}}]{Beach_2009}
\bibinfo{author}{\bibfnamefont{K.~S.~D.} \bibnamefont{Beach}},
  \bibinfo{author}{\bibfnamefont{F.}~\bibnamefont{Alet}},
  \bibinfo{author}{\bibfnamefont{M.}~\bibnamefont{Mambrini}}, \bibnamefont{and}
  \bibinfo{author}{\bibfnamefont{S.}~\bibnamefont{Capponi}},
  \bibinfo{journal}{Phys. Rev. B} \textbf{\bibinfo{volume}{80}},
  \bibinfo{pages}{184401} (\bibinfo{year}{2009}),
  \urlprefix\url{https://link.aps.org/doi/10.1103/PhysRevB.80.184401}.

\bibitem[{\citenamefont{Evertz et~al.}(1993)\citenamefont{Evertz, Lana, and
  Marcu}}]{Evertz_1993}
\bibinfo{author}{\bibfnamefont{H.~G.} \bibnamefont{Evertz}},
  \bibinfo{author}{\bibfnamefont{G.}~\bibnamefont{Lana}}, \bibnamefont{and}
  \bibinfo{author}{\bibfnamefont{M.}~\bibnamefont{Marcu}},
  \bibinfo{journal}{Phys. Rev. Lett.} \textbf{\bibinfo{volume}{70}},
  \bibinfo{pages}{875} (\bibinfo{year}{1993}),
  \urlprefix\url{https://link.aps.org/doi/10.1103/PhysRevLett.70.875}.

\bibitem[{\citenamefont{Ying et~al.}(1993)\citenamefont{Ying, Wiese, and
  Ji}}]{Ying_1993}
\bibinfo{author}{\bibfnamefont{H.-P.} \bibnamefont{Ying}},
  \bibinfo{author}{\bibfnamefont{U.-J.} \bibnamefont{Wiese}}, \bibnamefont{and}
  \bibinfo{author}{\bibfnamefont{D.-R.} \bibnamefont{Ji}},
  \bibinfo{journal}{Physics Letters A} \textbf{\bibinfo{volume}{183}},
  \bibinfo{pages}{441} (\bibinfo{year}{1993}), ISSN \bibinfo{issn}{0375-9601},
  \urlprefix\url{https://www.sciencedirect.com/science/article/pii/037596019390603W}.

\bibitem[{\citenamefont{Kawashima and Gubernatis}(1994)}]{Kawashima_1994}
\bibinfo{author}{\bibfnamefont{N.}~\bibnamefont{Kawashima}} \bibnamefont{and}
  \bibinfo{author}{\bibfnamefont{J.~E.} \bibnamefont{Gubernatis}},
  \bibinfo{journal}{Phys. Rev. Lett.} \textbf{\bibinfo{volume}{73}},
  \bibinfo{pages}{1295} (\bibinfo{year}{1994}),
  \urlprefix\url{https://link.aps.org/doi/10.1103/PhysRevLett.73.1295}.

\bibitem[{\citenamefont{Sandvik}(1999)}]{Sandvik_1999}
\bibinfo{author}{\bibfnamefont{A.~W.} \bibnamefont{Sandvik}},
  \bibinfo{journal}{Phys. Rev. B} \textbf{\bibinfo{volume}{59}},
  \bibinfo{pages}{R14157} (\bibinfo{year}{1999}),
  \urlprefix\url{https://link.aps.org/doi/10.1103/PhysRevB.59.R14157}.

\bibitem[{\citenamefont{Evertz}(2003)}]{Evertz_loop_review}
\bibinfo{author}{\bibfnamefont{H.~G.} \bibnamefont{Evertz}},
  \bibinfo{journal}{Advances in Physics} \textbf{\bibinfo{volume}{52}},
  \bibinfo{pages}{1} (\bibinfo{year}{2003}),
  \eprint{https://doi.org/10.1080/0001873021000049195},
  \urlprefix\url{https://doi.org/10.1080/0001873021000049195}.

\bibitem[{\citenamefont{Sandvik and Evertz}(2010)}]{Sandvik_2010b}
\bibinfo{author}{\bibfnamefont{A.~W.} \bibnamefont{Sandvik}} \bibnamefont{and}
  \bibinfo{author}{\bibfnamefont{H.~G.} \bibnamefont{Evertz}},
  \bibinfo{journal}{Phys. Rev. B} \textbf{\bibinfo{volume}{82}},
  \bibinfo{pages}{024407} (\bibinfo{year}{2010}),
  \urlprefix\url{https://link.aps.org/doi/10.1103/PhysRevB.82.024407}.

\bibitem[{\citenamefont{Prokof'ev et~al.}(1998)\citenamefont{Prokof'ev,
  Svistunov, and Tupitsyn}}]{Prokofev_1998}
\bibinfo{author}{\bibfnamefont{N.~V.} \bibnamefont{Prokof'ev}},
  \bibinfo{author}{\bibfnamefont{B.~V.} \bibnamefont{Svistunov}},
  \bibnamefont{and} \bibinfo{author}{\bibfnamefont{I.~S.}
  \bibnamefont{Tupitsyn}}, \bibinfo{journal}{Journal of Experimental and
  Theoretical Physics} \textbf{\bibinfo{volume}{87}}, \bibinfo{pages}{310}
  (\bibinfo{year}{1998}), ISSN \bibinfo{issn}{1090-6509},
  \urlprefix\url{https://doi.org/10.1134/1.558661}.

\bibitem[{\citenamefont{Sylju\aa{}sen and
  Sandvik}(2002)}]{Syljuasen_Sandvik_2002}
\bibinfo{author}{\bibfnamefont{O.~F.} \bibnamefont{Sylju\aa{}sen}}
  \bibnamefont{and} \bibinfo{author}{\bibfnamefont{A.~W.}
  \bibnamefont{Sandvik}}, \bibinfo{journal}{Phys. Rev. E}
  \textbf{\bibinfo{volume}{66}}, \bibinfo{pages}{046701}
  (\bibinfo{year}{2002}),
  \urlprefix\url{https://link.aps.org/doi/10.1103/PhysRevE.66.046701}.

\bibitem[{\citenamefont{Sandvik}(2003)}]{Sandvik_2003}
\bibinfo{author}{\bibfnamefont{A.~W.} \bibnamefont{Sandvik}},
  \bibinfo{journal}{Phys. Rev. E} \textbf{\bibinfo{volume}{68}},
  \bibinfo{pages}{056701} (\bibinfo{year}{2003}),
  \urlprefix\url{https://link.aps.org/doi/10.1103/PhysRevE.68.056701}.

\bibitem[{\citenamefont{Alet et~al.}(2005)\citenamefont{Alet, Wessel, and
  Troyer}}]{Alet_Wessel_Troyer_2005}
\bibinfo{author}{\bibfnamefont{F.}~\bibnamefont{Alet}},
  \bibinfo{author}{\bibfnamefont{S.}~\bibnamefont{Wessel}}, \bibnamefont{and}
  \bibinfo{author}{\bibfnamefont{M.}~\bibnamefont{Troyer}},
  \bibinfo{journal}{Phys. Rev. E} \textbf{\bibinfo{volume}{71}},
  \bibinfo{pages}{036706} (\bibinfo{year}{2005}),
  \urlprefix\url{https://link.aps.org/doi/10.1103/PhysRevE.71.036706}.

\bibitem[{\citenamefont{Heidarian and Damle}(2005)}]{Heidarian_2005}
\bibinfo{author}{\bibfnamefont{D.}~\bibnamefont{Heidarian}} \bibnamefont{and}
  \bibinfo{author}{\bibfnamefont{K.}~\bibnamefont{Damle}},
  \bibinfo{journal}{Phys. Rev. Lett.} \textbf{\bibinfo{volume}{95}},
  \bibinfo{pages}{127206} (\bibinfo{year}{2005}),
  \urlprefix\url{https://link.aps.org/doi/10.1103/PhysRevLett.95.127206}.

\bibitem[{\citenamefont{Isakov et~al.}(2006)\citenamefont{Isakov, Wessel,
  Melko, Sengupta, and Kim}}]{Isakov_2006}
\bibinfo{author}{\bibfnamefont{S.~V.} \bibnamefont{Isakov}},
  \bibinfo{author}{\bibfnamefont{S.}~\bibnamefont{Wessel}},
  \bibinfo{author}{\bibfnamefont{R.~G.} \bibnamefont{Melko}},
  \bibinfo{author}{\bibfnamefont{K.}~\bibnamefont{Sengupta}}, \bibnamefont{and}
  \bibinfo{author}{\bibfnamefont{Y.~B.} \bibnamefont{Kim}},
  \bibinfo{journal}{Phys. Rev. Lett.} \textbf{\bibinfo{volume}{97}},
  \bibinfo{pages}{147202} (\bibinfo{year}{2006}),
  \urlprefix\url{https://link.aps.org/doi/10.1103/PhysRevLett.97.147202}.

\bibitem[{\citenamefont{Melko}(2007)}]{Melko_2007}
\bibinfo{author}{\bibfnamefont{R.~G.} \bibnamefont{Melko}},
  \bibinfo{journal}{Journal of Physics: Condensed Matter}
  \textbf{\bibinfo{volume}{19}}, \bibinfo{pages}{145203}
  (\bibinfo{year}{2007}),
  \urlprefix\url{https://doi.org/10.1088/0953-8984/19/14/145203}.

\bibitem[{\citenamefont{Biswas et~al.}(2016)\citenamefont{Biswas, Rakala, and
  Damle}}]{Biswas_2016}
\bibinfo{author}{\bibfnamefont{S.}~\bibnamefont{Biswas}},
  \bibinfo{author}{\bibfnamefont{G.}~\bibnamefont{Rakala}}, \bibnamefont{and}
  \bibinfo{author}{\bibfnamefont{K.}~\bibnamefont{Damle}},
  \bibinfo{journal}{Phys. Rev. B} \textbf{\bibinfo{volume}{93}},
  \bibinfo{pages}{235103} (\bibinfo{year}{2016}),
  \urlprefix\url{https://link.aps.org/doi/10.1103/PhysRevB.93.235103}.

\bibitem[{\citenamefont{Swendsen and Wang}(1987)}]{Swendsen_Wang_1987}
\bibinfo{author}{\bibfnamefont{R.~H.} \bibnamefont{Swendsen}} \bibnamefont{and}
  \bibinfo{author}{\bibfnamefont{J.-S.} \bibnamefont{Wang}},
  \bibinfo{journal}{Phys. Rev. Lett.} \textbf{\bibinfo{volume}{58}},
  \bibinfo{pages}{86} (\bibinfo{year}{1987}),
  \urlprefix\url{https://link.aps.org/doi/10.1103/PhysRevLett.58.86}.

\bibitem[{\citenamefont{Wolff}(1989)}]{Wolff_1989}
\bibinfo{author}{\bibfnamefont{U.}~\bibnamefont{Wolff}},
  \bibinfo{journal}{Phys. Rev. Lett.} \textbf{\bibinfo{volume}{62}},
  \bibinfo{pages}{361} (\bibinfo{year}{1989}),
  \urlprefix\url{https://link.aps.org/doi/10.1103/PhysRevLett.62.361}.

\bibitem[{\citenamefont{Kaul}(2015)}]{Kaul_2015a}
\bibinfo{author}{\bibfnamefont{R.~K.} \bibnamefont{Kaul}},
  \bibinfo{journal}{Phys. Rev. B} \textbf{\bibinfo{volume}{91}},
  \bibinfo{pages}{054413} (\bibinfo{year}{2015}),
  \urlprefix\url{https://link.aps.org/doi/10.1103/PhysRevB.91.054413}.

\bibitem[{Bea()}]{Beach_footnote}
\bibinfo{note}{See footnote [19] of Beach \textit{et al}~\cite{Beach_2009}, and
  the associated text.}

\bibitem[{\citenamefont{Nahum et~al.}(2011)\citenamefont{Nahum, Chalker, Serna,
  Ortu\~no, and Somoza}}]{Nahum_etal_2011}
\bibinfo{author}{\bibfnamefont{A.}~\bibnamefont{Nahum}},
  \bibinfo{author}{\bibfnamefont{J.~T.} \bibnamefont{Chalker}},
  \bibinfo{author}{\bibfnamefont{P.}~\bibnamefont{Serna}},
  \bibinfo{author}{\bibfnamefont{M.}~\bibnamefont{Ortu\~no}}, \bibnamefont{and}
  \bibinfo{author}{\bibfnamefont{A.~M.} \bibnamefont{Somoza}},
  \bibinfo{journal}{Phys. Rev. Lett.} \textbf{\bibinfo{volume}{107}},
  \bibinfo{pages}{110601} (\bibinfo{year}{2011}),
  \urlprefix\url{https://link.aps.org/doi/10.1103/PhysRevLett.107.110601}.

\bibitem[{\citenamefont{Nahum et~al.}(2013)\citenamefont{Nahum, Chalker, Serna,
  Ortu\~no, and Somoza}}]{Nahum_etal_2013}
\bibinfo{author}{\bibfnamefont{A.}~\bibnamefont{Nahum}},
  \bibinfo{author}{\bibfnamefont{J.~T.} \bibnamefont{Chalker}},
  \bibinfo{author}{\bibfnamefont{P.}~\bibnamefont{Serna}},
  \bibinfo{author}{\bibfnamefont{M.}~\bibnamefont{Ortu\~no}}, \bibnamefont{and}
  \bibinfo{author}{\bibfnamefont{A.~M.} \bibnamefont{Somoza}},
  \bibinfo{journal}{Phys. Rev. B} \textbf{\bibinfo{volume}{88}},
  \bibinfo{pages}{134411} (\bibinfo{year}{2013}),
  \urlprefix\url{https://link.aps.org/doi/10.1103/PhysRevB.88.134411}.

\bibitem[{\citenamefont{Nahum et~al.}(2015)\citenamefont{Nahum, Chalker, Serna,
  Ortu\~no, and Somoza}}]{Nahum_etal_2015}
\bibinfo{author}{\bibfnamefont{A.}~\bibnamefont{Nahum}},
  \bibinfo{author}{\bibfnamefont{J.~T.} \bibnamefont{Chalker}},
  \bibinfo{author}{\bibfnamefont{P.}~\bibnamefont{Serna}},
  \bibinfo{author}{\bibfnamefont{M.}~\bibnamefont{Ortu\~no}}, \bibnamefont{and}
  \bibinfo{author}{\bibfnamefont{A.~M.} \bibnamefont{Somoza}},
  \bibinfo{journal}{Phys. Rev. X} \textbf{\bibinfo{volume}{5}},
  \bibinfo{pages}{041048} (\bibinfo{year}{2015}),
  \urlprefix\url{https://link.aps.org/doi/10.1103/PhysRevX.5.041048}.

\bibitem[{\citenamefont{Aizenman and Nachtergaele}(1994)}]{Aizenman_1994}
\bibinfo{author}{\bibfnamefont{M.}~\bibnamefont{Aizenman}} \bibnamefont{and}
  \bibinfo{author}{\bibfnamefont{B.}~\bibnamefont{Nachtergaele}},
  \bibinfo{journal}{Communications in Mathematical Physics}
  \textbf{\bibinfo{volume}{164}}, \bibinfo{pages}{17} (\bibinfo{year}{1994}),
  ISSN \bibinfo{issn}{1432-0916},
  \urlprefix\url{https://doi.org/10.1007/BF02108805}.

\bibitem[{\citenamefont{Kawashima and Gubernatis}(1995)}]{Kawashima_1995}
\bibinfo{author}{\bibfnamefont{N.}~\bibnamefont{Kawashima}} \bibnamefont{and}
  \bibinfo{author}{\bibfnamefont{J.~E.} \bibnamefont{Gubernatis}},
  \bibinfo{journal}{Journal of Statistical Physics}
  \textbf{\bibinfo{volume}{80}}, \bibinfo{pages}{169} (\bibinfo{year}{1995}),
  ISSN \bibinfo{issn}{1572-9613},
  \urlprefix\url{https://doi.org/10.1007/BF02178358}.

\bibitem[{\citenamefont{Suzuki}(1976)}]{Suzuki_1976}
\bibinfo{author}{\bibfnamefont{M.}~\bibnamefont{Suzuki}},
  \bibinfo{journal}{Progress of Theoretical Physics}
  \textbf{\bibinfo{volume}{56}}, \bibinfo{pages}{1454} (\bibinfo{year}{1976}),
  ISSN \bibinfo{issn}{0033-068X},
  \eprint{https://academic.oup.com/ptp/article-pdf/56/5/1454/5264429/56-5-1454.pdf},
  \urlprefix\url{https://doi.org/10.1143/PTP.56.1454}.

\bibitem[{\citenamefont{Desai and Kaul}(2019)}]{Desai_2019}
\bibinfo{author}{\bibfnamefont{N.}~\bibnamefont{Desai}} \bibnamefont{and}
  \bibinfo{author}{\bibfnamefont{R.~K.} \bibnamefont{Kaul}},
  \bibinfo{journal}{Phys. Rev. Lett.} \textbf{\bibinfo{volume}{123}},
  \bibinfo{pages}{107202} (\bibinfo{year}{2019}),
  \urlprefix\url{https://link.aps.org/doi/10.1103/PhysRevLett.123.107202}.

\bibitem[{\citenamefont{Desai}(2020)}]{ND_thesis}
\bibinfo{author}{\bibfnamefont{N.}~\bibnamefont{Desai}}, Ph.D. thesis,
  \bibinfo{school}{University of Kentucky}, \bibinfo{address}{Lexington, KY,
  USA} (\bibinfo{year}{2020}).

\bibitem[{\citenamefont{Kaul}(2011)}]{Kaul_2011}
\bibinfo{author}{\bibfnamefont{R.~K.} \bibnamefont{Kaul}},
  \bibinfo{journal}{Phys. Rev. B} \textbf{\bibinfo{volume}{84}},
  \bibinfo{pages}{054407} (\bibinfo{year}{2011}),
  \urlprefix\url{https://link.aps.org/doi/10.1103/PhysRevB.84.054407}.

\bibitem[{ant()}]{antifundamental_footnote}
\bibinfo{note}{Formally, for $N>2$, the magnetic degrees of freedom are in the
  fundamental representation on one sublattice, while they are in the
  conjugate-to-fundamental representation on the other sublattice. See e.g.
  Ref.~\onlinecite{Kaul_2011}}.

\bibitem[{sup()}]{supp}
\bibinfo{note}{See supplementary document for further details on the
  implementation of the resummation-based updates, and some other technical
  details. Also, the final section discusses the approach to the quantum dimer
  model Hilbert space in the large-$N$ regime for the uncolored loop
  representation.}

\bibitem[{\citenamefont{Harada et~al.}(2003)\citenamefont{Harada, Kawashima,
  and Troyer}}]{kawashima_harada_troyer_prl2003}
\bibinfo{author}{\bibfnamefont{K.}~\bibnamefont{Harada}},
  \bibinfo{author}{\bibfnamefont{N.}~\bibnamefont{Kawashima}},
  \bibnamefont{and} \bibinfo{author}{\bibfnamefont{M.}~\bibnamefont{Troyer}},
  \bibinfo{journal}{Phys. Rev. Lett.} \textbf{\bibinfo{volume}{90}},
  \bibinfo{pages}{117203} (\bibinfo{year}{2003}),
  \urlprefix\url{https://link.aps.org/doi/10.1103/PhysRevLett.90.117203}.

\bibitem[{\citenamefont{Affleck et~al.}(1988)\citenamefont{Affleck, Kennedy,
  Lieb, and Tasaki}}]{Affleck_1988}
\bibinfo{author}{\bibfnamefont{I.}~\bibnamefont{Affleck}},
  \bibinfo{author}{\bibfnamefont{T.}~\bibnamefont{Kennedy}},
  \bibinfo{author}{\bibfnamefont{E.~H.} \bibnamefont{Lieb}}, \bibnamefont{and}
  \bibinfo{author}{\bibfnamefont{H.}~\bibnamefont{Tasaki}},
  \bibinfo{journal}{Communications in Mathematical Physics}
  \textbf{\bibinfo{volume}{115}}, \bibinfo{pages}{477} (\bibinfo{year}{1988}),
  ISSN \bibinfo{issn}{1432-0916},
  \urlprefix\url{https://doi.org/10.1007/BF01218021}.

\bibitem[{err()}]{error_footnote}
\bibinfo{note}{The errors on the data points are smaller in size than the
  symbols representing them, and are hence not explicitly visible in the
  plots.}

\bibitem[{\citenamefont{Yan et~al.}(2019)\citenamefont{Yan, Wu, Liu,
  Sylju\aa{}sen, Lou, and Chen}}]{sweeping_cluster_2019}
\bibinfo{author}{\bibfnamefont{Z.}~\bibnamefont{Yan}},
  \bibinfo{author}{\bibfnamefont{Y.}~\bibnamefont{Wu}},
  \bibinfo{author}{\bibfnamefont{C.}~\bibnamefont{Liu}},
  \bibinfo{author}{\bibfnamefont{O.~F.} \bibnamefont{Sylju\aa{}sen}},
  \bibinfo{author}{\bibfnamefont{J.}~\bibnamefont{Lou}}, \bibnamefont{and}
  \bibinfo{author}{\bibfnamefont{Y.}~\bibnamefont{Chen}},
  \bibinfo{journal}{Phys. Rev. B} \textbf{\bibinfo{volume}{99}},
  \bibinfo{pages}{165135} (\bibinfo{year}{2019}),
  \urlprefix\url{https://link.aps.org/doi/10.1103/PhysRevB.99.165135}.

\bibitem[{\citenamefont{Yan et~al.}(2021)\citenamefont{Yan, Zhou,
  Sylju\aa{}sen, Zhang, Yuan, Lou, and Chen}}]{Yan_etal_2021}
\bibinfo{author}{\bibfnamefont{Z.}~\bibnamefont{Yan}},
  \bibinfo{author}{\bibfnamefont{Z.}~\bibnamefont{Zhou}},
  \bibinfo{author}{\bibfnamefont{O.~F.} \bibnamefont{Sylju\aa{}sen}},
  \bibinfo{author}{\bibfnamefont{J.}~\bibnamefont{Zhang}},
  \bibinfo{author}{\bibfnamefont{T.}~\bibnamefont{Yuan}},
  \bibinfo{author}{\bibfnamefont{J.}~\bibnamefont{Lou}}, \bibnamefont{and}
  \bibinfo{author}{\bibfnamefont{Y.}~\bibnamefont{Chen}},
  \bibinfo{journal}{Phys. Rev. B} \textbf{\bibinfo{volume}{103}},
  \bibinfo{pages}{094421} (\bibinfo{year}{2021}),
  \urlprefix\url{https://link.aps.org/doi/10.1103/PhysRevB.103.094421}.

\bibitem[{\citenamefont{Todo and Kato}(2001)}]{Todo_2001}
\bibinfo{author}{\bibfnamefont{S.}~\bibnamefont{Todo}} \bibnamefont{and}
  \bibinfo{author}{\bibfnamefont{K.}~\bibnamefont{Kato}},
  \bibinfo{journal}{Phys. Rev. Lett.} \textbf{\bibinfo{volume}{87}},
  \bibinfo{pages}{047203} (\bibinfo{year}{2001}),
  \urlprefix\url{https://link.aps.org/doi/10.1103/PhysRevLett.87.047203}.

\bibitem[{\citenamefont{Kawashima and Harada}(2004)}]{Kawashima_2004}
\bibinfo{author}{\bibfnamefont{N.}~\bibnamefont{Kawashima}} \bibnamefont{and}
  \bibinfo{author}{\bibfnamefont{K.}~\bibnamefont{Harada}},
  \bibinfo{journal}{Journal of the Physical Society of Japan}
  \textbf{\bibinfo{volume}{73}}, \bibinfo{pages}{1379} (\bibinfo{year}{2004}),
  \eprint{https://doi.org/10.1143/JPSJ.73.1379},
  \urlprefix\url{https://doi.org/10.1143/JPSJ.73.1379}.

\bibitem[{foo()}]{footnote_ND}
\bibinfo{note}{See supplementary of Ref.~\onlinecite{Desai_2019}.}

\bibitem[{\citenamefont{Kawashima and Tanabe}(2007)}]{kawashima_tanabe_prl2007}
\bibinfo{author}{\bibfnamefont{N.}~\bibnamefont{Kawashima}} \bibnamefont{and}
  \bibinfo{author}{\bibfnamefont{Y.}~\bibnamefont{Tanabe}},
  \bibinfo{journal}{Phys. Rev. Lett.} \textbf{\bibinfo{volume}{98}},
  \bibinfo{pages}{057202} (\bibinfo{year}{2007}),
  \urlprefix\url{https://link.aps.org/doi/10.1103/PhysRevLett.98.057202}.

\bibitem[{\citenamefont{Okubo et~al.}(2015)\citenamefont{Okubo, Harada, Lou,
  and Kawashima}}]{okubo_etal_prb2015}
\bibinfo{author}{\bibfnamefont{T.}~\bibnamefont{Okubo}},
  \bibinfo{author}{\bibfnamefont{K.}~\bibnamefont{Harada}},
  \bibinfo{author}{\bibfnamefont{J.}~\bibnamefont{Lou}}, \bibnamefont{and}
  \bibinfo{author}{\bibfnamefont{N.}~\bibnamefont{Kawashima}},
  \bibinfo{journal}{Phys. Rev. B} \textbf{\bibinfo{volume}{92}},
  \bibinfo{pages}{134404} (\bibinfo{year}{2015}),
  \urlprefix\url{https://link.aps.org/doi/10.1103/PhysRevB.92.134404}.

\bibitem[{pro()}]{proj_footnote}
\bibinfo{note}{Zero temperature projector variants of the resummed SSE
  algorithm in higher symmetric representations can also now be envisaged.}

\bibitem[{\citenamefont{Beach and Sandvik}(2006)}]{Beach_2006}
\bibinfo{author}{\bibfnamefont{K.}~\bibnamefont{Beach}} \bibnamefont{and}
  \bibinfo{author}{\bibfnamefont{A.~W.} \bibnamefont{Sandvik}},
  \bibinfo{journal}{Nuclear Physics B} \textbf{\bibinfo{volume}{750}},
  \bibinfo{pages}{142} (\bibinfo{year}{2006}), ISSN \bibinfo{issn}{0550-3213},
  \urlprefix\url{https://www.sciencedirect.com/science/article/pii/S0550321306004214}.

\bibitem[{Nah({\natexlab{a}})}]{Nahum_PRB_IIIA}
\bibinfo{note}{See in particular Sec. IIIA of
  Ref.~\onlinecite{Nahum_etal_2013}}.

\bibitem[{Nah({\natexlab{b}})}]{Nahum_PRB_IIID}
\bibinfo{note}{See in particular Sec. IIID of
  Ref.~\onlinecite{Nahum_etal_2013}}.

\bibitem[{\citenamefont{Sandvik}(2007)}]{Sandvik_2007}
\bibinfo{author}{\bibfnamefont{A.~W.} \bibnamefont{Sandvik}},
  \bibinfo{journal}{Phys. Rev. Lett.} \textbf{\bibinfo{volume}{98}},
  \bibinfo{pages}{227202} (\bibinfo{year}{2007}),
  \urlprefix\url{https://link.aps.org/doi/10.1103/PhysRevLett.98.227202}.

\bibitem[{\citenamefont{Senthil
  et~al.}(2004{\natexlab{a}})\citenamefont{Senthil, Vishwanath, Balents,
  Sachdev, and Fisher}}]{Senthil_etal_2004a}
\bibinfo{author}{\bibfnamefont{T.}~\bibnamefont{Senthil}},
  \bibinfo{author}{\bibfnamefont{A.}~\bibnamefont{Vishwanath}},
  \bibinfo{author}{\bibfnamefont{L.}~\bibnamefont{Balents}},
  \bibinfo{author}{\bibfnamefont{S.}~\bibnamefont{Sachdev}}, \bibnamefont{and}
  \bibinfo{author}{\bibfnamefont{M.~P.~A.} \bibnamefont{Fisher}},
  \bibinfo{journal}{Science} \textbf{\bibinfo{volume}{303}},
  \bibinfo{pages}{1490} (\bibinfo{year}{2004}{\natexlab{a}}), ISSN
  \bibinfo{issn}{0036-8075},
  \eprint{https://science.sciencemag.org/content/303/5663/1490.full.pdf},
  \urlprefix\url{https://science.sciencemag.org/content/303/5663/1490}.

\bibitem[{\citenamefont{Senthil
  et~al.}(2004{\natexlab{b}})\citenamefont{Senthil, Balents, Sachdev,
  Vishwanath, and Fisher}}]{Senthil_etal_2004b}
\bibinfo{author}{\bibfnamefont{T.}~\bibnamefont{Senthil}},
  \bibinfo{author}{\bibfnamefont{L.}~\bibnamefont{Balents}},
  \bibinfo{author}{\bibfnamefont{S.}~\bibnamefont{Sachdev}},
  \bibinfo{author}{\bibfnamefont{A.}~\bibnamefont{Vishwanath}},
  \bibnamefont{and} \bibinfo{author}{\bibfnamefont{M.~P.~A.}
  \bibnamefont{Fisher}}, \bibinfo{journal}{Phys. Rev. B}
  \textbf{\bibinfo{volume}{70}}, \bibinfo{pages}{144407}
  (\bibinfo{year}{2004}{\natexlab{b}}),
  \urlprefix\url{https://link.aps.org/doi/10.1103/PhysRevB.70.144407}.

\bibitem[{Nah({\natexlab{c}})}]{Nahum_PRX_lattice}
\bibinfo{note}{In Ref.~\onlinecite{Nahum_etal_2015}, the underlying lattice of
  the classical loop gas is the so-called ``3D L lattice'' (Fig. 1 of
  Ref.~\onlinecite{Nahum_etal_2015}), while the $JQ$ model on square lattice
  maps to a loop gas on a simple cubic lattice. Such a difference is expected
  to not affect the long-distance universal physics, since both lattice have
  the same cubic symmetry. Same goes for the microscopic differences in the
  transfer matrices.}

\end{thebibliography}


\begin{thebibliography}{6}%
\makeatletter
\providecommand \@ifxundefined [1]{%
 \@ifx{#1\undefined}
}%
\providecommand \@ifnum [1]{%
 \ifnum #1\expandafter \@firstoftwo
 \else \expandafter \@secondoftwo
 \fi
}%
\providecommand \@ifx [1]{%
 \ifx #1\expandafter \@firstoftwo
 \else \expandafter \@secondoftwo
 \fi
}%
\providecommand \natexlab [1]{#1}%
\providecommand \enquote  [1]{``#1''}%
\providecommand \bibnamefont  [1]{#1}%
\providecommand \bibfnamefont [1]{#1}%
\providecommand \citenamefont [1]{#1}%
\providecommand \href@noop [0]{\@secondoftwo}%
\providecommand \href [0]{\begingroup \@sanitize@url \@href}%
\providecommand \@href[1]{\@@startlink{#1}\@@href}%
\providecommand \@@href[1]{\endgroup#1\@@endlink}%
\providecommand \@sanitize@url [0]{\catcode `\\12\catcode `\$12\catcode
  `\&12\catcode `\#12\catcode `\^12\catcode `\_12\catcode `\%12\relax}%
\providecommand \@@startlink[1]{}%
\providecommand \@@endlink[0]{}%
\providecommand \url  [0]{\begingroup\@sanitize@url \@url }%
\providecommand \@url [1]{\endgroup\@href {#1}{\urlprefix }}%
\providecommand \urlprefix  [0]{URL }%
\providecommand \Eprint [0]{\href }%
\providecommand \doibase [0]{http://dx.doi.org/}%
\providecommand \selectlanguage [0]{\@gobble}%
\providecommand \bibinfo  [0]{\@secondoftwo}%
\providecommand \bibfield  [0]{\@secondoftwo}%
\providecommand \translation [1]{[#1]}%
\providecommand \BibitemOpen [0]{}%
\providecommand \bibitemStop [0]{}%
\providecommand \bibitemNoStop [0]{.\EOS\space}%
\providecommand \EOS [0]{\spacefactor3000\relax}%
\providecommand \BibitemShut  [1]{\csname bibitem#1\endcsname}%
\let\auto@bib@innerbib\@empty
\bibitem [{\citenamefont {{Sandvik}}(2010)}]{Sandvik_2010_review}%
  \BibitemOpen
  \bibfield  {author} {\bibinfo {author} {\bibfnamefont {A.~W.}\ \bibnamefont
  {{Sandvik}}},\ }in\ \href {\doibase 10.1063/1.3518900} {\emph {\bibinfo
  {booktitle} {Lectures on the Physics of Strongly Correlated Systems Xiv:
  Fourteenth Training Course in the Physics of Strongly Correlated Systems}}},\
  \bibinfo {series} {American Institute of Physics Conference Series}, Vol.\
  \bibinfo {volume} {1297},\ \bibinfo {editor} {edited by\ \bibinfo {editor}
  {\bibfnamefont {A.}~\bibnamefont {{Avella}}}\ and\ \bibinfo {editor}
  {\bibfnamefont {F.}~\bibnamefont {{Mancini}}}}\ (\bibinfo {year} {2010})\
  pp.\ \bibinfo {pages} {135--338},\ \Eprint {http://arxiv.org/abs/1101.3281}
  {arXiv:1101.3281 [cond-mat.str-el]} \BibitemShut {NoStop}%
\bibitem [{Note1()}]{Note1}%
  \BibitemOpen
  \bibinfo {note} {Our operator vertex has four legs since we are working with
  a two spin interaction. For other interactions with more number of spins,
  e.g. the $Q$-term, the full operator vertex can be treated as multiple four
  legged operator vertices at the same time slice.}\BibitemShut {Stop}%
\bibitem [{\citenamefont {Sandvik}(1997)}]{sandvik1997:prb}%
  \BibitemOpen
  \bibfield  {author} {\bibinfo {author} {\bibfnamefont {A.~W.}\ \bibnamefont
  {Sandvik}},\ }\href {\doibase 10.1103/PhysRevB.56.11678} {\bibfield
  {journal} {\bibinfo  {journal} {Phys. Rev. B}\ }\textbf {\bibinfo {volume}
  {56}},\ \bibinfo {pages} {11678} (\bibinfo {year} {1997})}\BibitemShut
  {NoStop}%
\bibitem [{\citenamefont {Sandvik}(2005)}]{Sandvik_2005}%
  \BibitemOpen
  \bibfield  {author} {\bibinfo {author} {\bibfnamefont {A.~W.}\ \bibnamefont
  {Sandvik}},\ }\href {\doibase 10.1103/PhysRevLett.95.207203} {\bibfield
  {journal} {\bibinfo  {journal} {Phys. Rev. Lett.}\ }\textbf {\bibinfo
  {volume} {95}},\ \bibinfo {pages} {207203} (\bibinfo {year}
  {2005})}\BibitemShut {NoStop}%
\bibitem [{\citenamefont {Sandvik}\ and\ \citenamefont
  {Evertz}(2010)}]{Sandvik_2010b}%
  \BibitemOpen
  \bibfield  {author} {\bibinfo {author} {\bibfnamefont {A.~W.}\ \bibnamefont
  {Sandvik}}\ and\ \bibinfo {author} {\bibfnamefont {H.~G.}\ \bibnamefont
  {Evertz}},\ }\href {\doibase 10.1103/PhysRevB.82.024407} {\bibfield
  {journal} {\bibinfo  {journal} {Phys. Rev. B}\ }\textbf {\bibinfo {volume}
  {82}},\ \bibinfo {pages} {024407} (\bibinfo {year} {2010})}\BibitemShut
  {NoStop}%
\bibitem [{\citenamefont {Beach}\ \emph {et~al.}(2009)\citenamefont {Beach},
  \citenamefont {Alet}, \citenamefont {Mambrini},\ and\ \citenamefont
  {Capponi}}]{Beach_2009}%
  \BibitemOpen
  \bibfield  {author} {\bibinfo {author} {\bibfnamefont {K.~S.~D.}\
  \bibnamefont {Beach}}, \bibinfo {author} {\bibfnamefont {F.}~\bibnamefont
  {Alet}}, \bibinfo {author} {\bibfnamefont {M.}~\bibnamefont {Mambrini}}, \
  and\ \bibinfo {author} {\bibfnamefont {S.}~\bibnamefont {Capponi}},\ }\href
  {\doibase 10.1103/PhysRevB.80.184401} {\bibfield  {journal} {\bibinfo
  {journal} {Phys. Rev. B}\ }\textbf {\bibinfo {volume} {80}},\ \bibinfo
  {pages} {184401} (\bibinfo {year} {2009})}\BibitemShut {NoStop}%
\end{thebibliography}%

\end{document}


\title{Supplemental Material}

\maketitle

\section{Details of the algorithm}
Our algorithm samples configurations of a closely-packed uncolored loop-gas ensemble as explained in the main text. For convenience, we adopt the standard step of working with operator strings of fixed-length, $M$, where the Hamiltonian operators are padded with 
identities (e.g. see Eqns. 258,263 of ~\cite{Sandvik_2010_review}).
Thus, the partition function can be re-written as
\begin{equation}
    Z(\beta)=\sum_{S_M} \frac{(M-n)!}{M!} (-\beta)^n N^{n_l}
     h_{b_1} h_{b_2} \ldots h_{b_M}
\end{equation}
where $n$ refers to the number of non-identity Hamiltonian operators,
$S_M$ is the fixed-length operator string with $n$ non-identities and
$M-n$ identities,
$h_{b_i}$ is either the spin-symmetric matrix element or an identity
matrix element, and the sum over $S_M$ implicitly includes the sum over
$n$ in the main text.
Therefore, for our example of $H=-J \sum_{\langle i,j \rangle} H_{ij}$ with
$H_{ij} = \frac{1}{N} \sum_\alpha |\alpha_i \alpha_j \rangle
\langle \alpha_i \alpha_j|$, the Monte Carlo weight of a configuration with $n$ number of 
non-identities and $n_l$ number of loops in the configuration is given by:
\begin{equation}
 W(n,M,n_l)=\bigg(\frac{\beta J}{N}\bigg)^n \frac{(M-n)!}{M!} N^{n_l}
\end{equation}
This is the weight of an operator string in standard fixed-length
SSE~\cite{Sandvik_2010_review} multiplied by the number of possible colorings of $n_l$ loops, $N^{n_l}$.

\begin{figure}[h]
\includegraphics[width=0.9\linewidth]{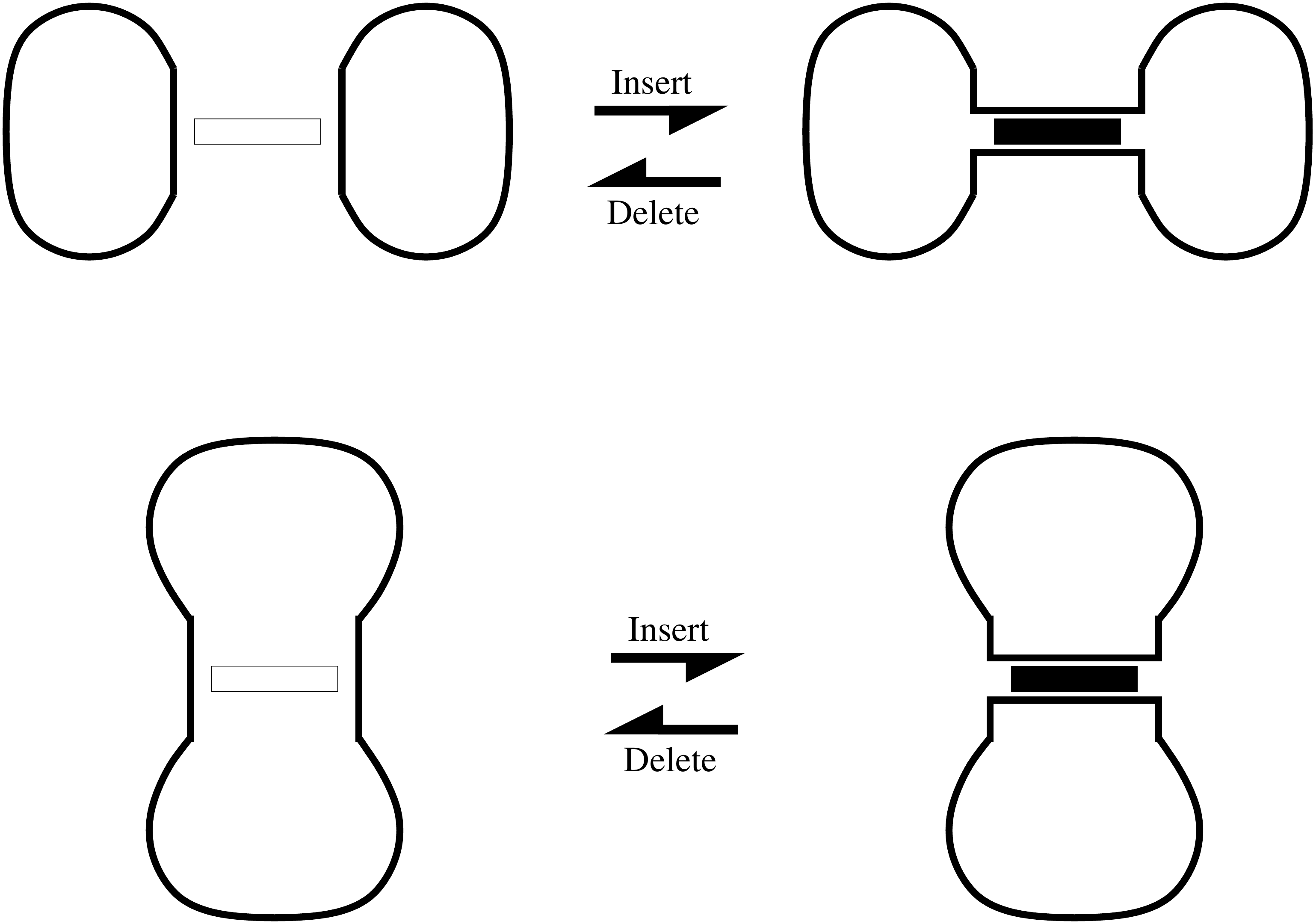}
\caption{ \label{suppfig:diag0}
Illustration of the basic update moves. The loop away from the
operator is not drawn in register with the discrete space-time lattice
which makes the point that the change in loop number $n_l$ during these
updates are purely determined by geometrical considerations.
}
\end{figure}

In our implementation of the algorithm, we propose the insertion of a non-identity 
operator on a random bond at a time slice that has an identity operator,
and the deletion of the (non-identity) operator if the time slice already has a non-identity
operator. The key step in the the update is the calculation of the change in 
number of loops due to insertion or deletion of an operator. 
The purely loop-topological aspect of this calculation is sketched in Fig.~\ref{suppfig:diag0}.
This is carried out algorithmically in the following way:
\begin{itemize}
  \item \textbf{Insertion}: If two space-time points are on the same loop, the insertion of a 
non-identity operator at these two points splits the loop into two ($n_l
\rightarrow n_l + 1 \implies \delta(n_l) = 1$). 
Conversely, if they are on different loops, the two loops merge into one on inserting 
an operator ($n_l \rightarrow n_l - 1 \implies \delta(n_l) = -1$). 
This has been shown schematically in Fig.~\ref{suppfig:diag0}. 
The Metropolis probability for acceptance of the insertion update: 
\begin{align}
P_{\text{accept}}(\text{identity}\rightarrow \text{non-identity}) & = \nonumber \\
\text{min}  [\frac{\beta J N_b}{N(M-n)} & N^{\delta(n_l)},1], 
\end{align}
where $N_b$ is the total number of bonds in the lattice.

  \item \textbf{Deletion}: If on an operator vertex, all the four legs
~\footnote{Our operator vertex has four legs since we are working with a two spin interaction. For other interactions with more number of spins, e.g. the $Q$-term, the full operator vertex can be treated as multiple four legged operator vertices at the same time slice.} 
of the operator vertex touch the same loop, the removal of the operator results in splitting of the loop into two ($n_l
\rightarrow n_l + 1 \implies \delta(n_l) = 1$). 
On the other hand if two different loops touch the two sides of the operator vertex, the deletion of the operator results in merging of the two loops 
($n_l \rightarrow n_l - 1 \implies \delta(n_l) = -1$).
This has been shown schematically in Fig.~\ref{suppfig:diag0}.
The Metropolis probability for acceptance of the insertion update is given by: 
\begin{align}
P_{\text{accept}}(\text{non-identity}\rightarrow \text{identity}) & = \nonumber \\
\text{min} [\frac{N(M-n+1)}{\beta J N_b } & N^{\delta n_l},1], 
\end{align}
where $N_b$ is the total number of bonds in the lattice. 
  
  \item \textbf{Rewiring}: In the case of more than one mini-spin for higher symmetric
representations, this update is carried out at the projection operator time slice. 
As shown in Fig 5 of the main manuscript, this kind of update can increase or 
decrease the number of loops. Therefore the update is accepted with the Metropolis probability 
of 
\begin{align}
P_{\text{accept}}(\text{rewire})=\text{min} [N^{\delta n_l},1]
\end{align}
  
\end{itemize}


In order to determine if two space-time points are on the same loop, we grow the loop starting from one of the points and check if the second point lies on that loop. 
We do this by storing the operator string as a linked-list data structure
which correctly stores the connections between the various non-identity operators
(as in Fig.~\ref{suppfig:diag1}, also see Fig. 57 of Ref.~\cite{Sandvik_2010_review}
and associated text for a discussion on this). 
This computation is time consuming in the long loop (magnetic) phases, but is very efficient in short-loop phases. We use two different implementations of the algorithm in the main manuscript:

\begin{figure}[t]
 \includegraphics[width=\linewidth]{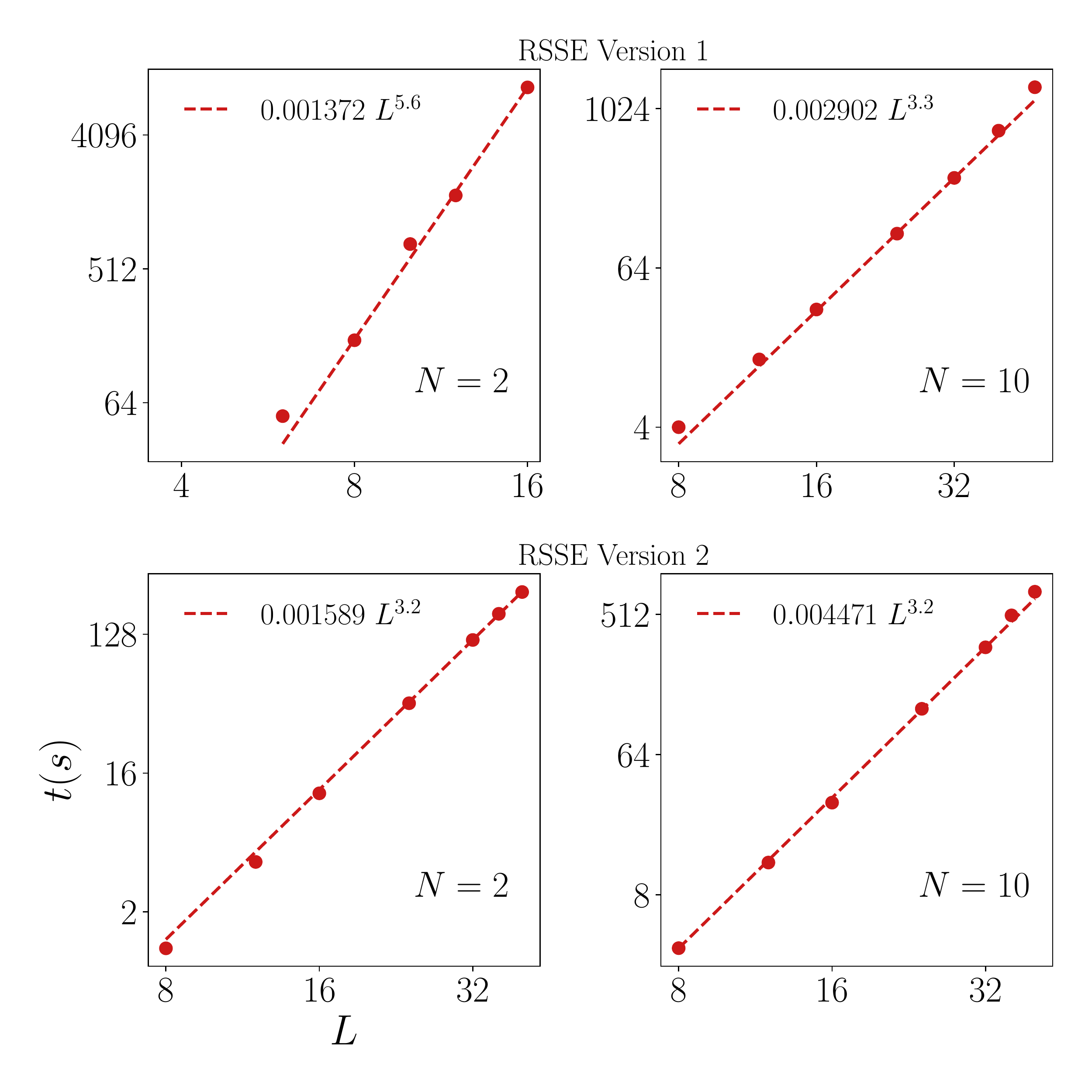}
 \caption{\label{suppfig:scaling} Time taken for $10^5$ Monte Carlo steps as a function of system size, $L$, with $\beta=L$ plotted on a log-log plot and fitted to a power-law for $N=2$ (long loop phase) and $N=10$ (short loop phase). One expects that operator string length, $M \propto L^2 \beta = L^3$. Version 1 of the resummed-SSE (RSSE) algorithm scales as O($M$) only in the short loop phase, while version 2 scales this way in both phases. (here version 1 and 2 are as explained in the text)
    }
\end{figure}

\begin{figure}[t]
 \includegraphics[width=0.9\linewidth]{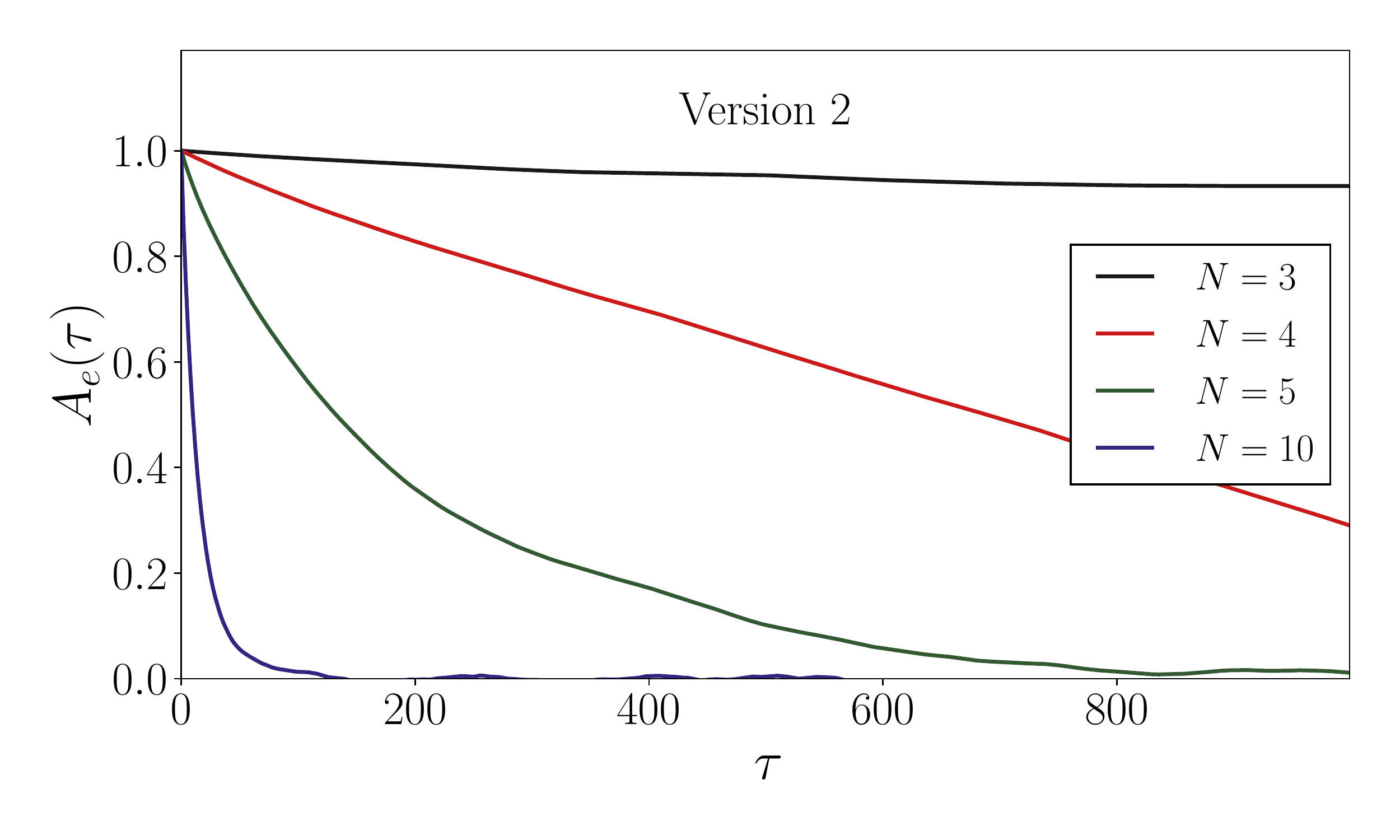}
 \caption{\label{enrautocorr_comparison_vsN} Autocorrelation of the energy per unit site on a $16 \times 16$ lattice measured after each Monte Carlo step using version 2 of our algorithm for $N=3,\, 4,\, 5,\, 10$ . The autocorrelations systematically get better as $N$ increases. }
\end{figure}

\begin{itemize}
 \item \emph{Version 1}: An insertion/deletion update is proposed at every time slice. Only in the short loop phase, does this version scale as O($M$), similar to standard SSE. In the long loop phase, however, one typically needs to grow loops of O($M$) length to carry out each update, resulting in costly and hence not as efficient updates. The contrast in performance of this algorithm in the two kinds phases is shown in Fig.~\ref{suppfig:scaling}.

 \item \emph{Version 2}: An update is proposed at a random time slice. The number of such updates per Monte Carlo step is determined by requiring that the number of sites touched while growing loops during each insertion/deletion operation is O($M$). Clearly, this update scales as O($M$) in either phase (see Fig.~\ref{suppfig:scaling}). In the long loop phases, there are much fewer updates per Monte Carlo step leading to a less efficient update compared to the short loop phase. This update progressively does better as one goes deep into the short loop phase as can be seen from the autocorrelations of energy in Fig.~\ref{enrautocorr_comparison_vsN}.
\end{itemize}

Among the two versions, we find that version 1 performs better when not too deep in the short loop phase. As the loops get shorter, the number of updates per MC step increase for version 2, thus performing better than version 1. The data for $N>15$ has been collected using version 2 of the algorithm.

\begin{figure}
 \includegraphics[width=\linewidth,clip=true,trim=0 12mm 0mm 0mm]{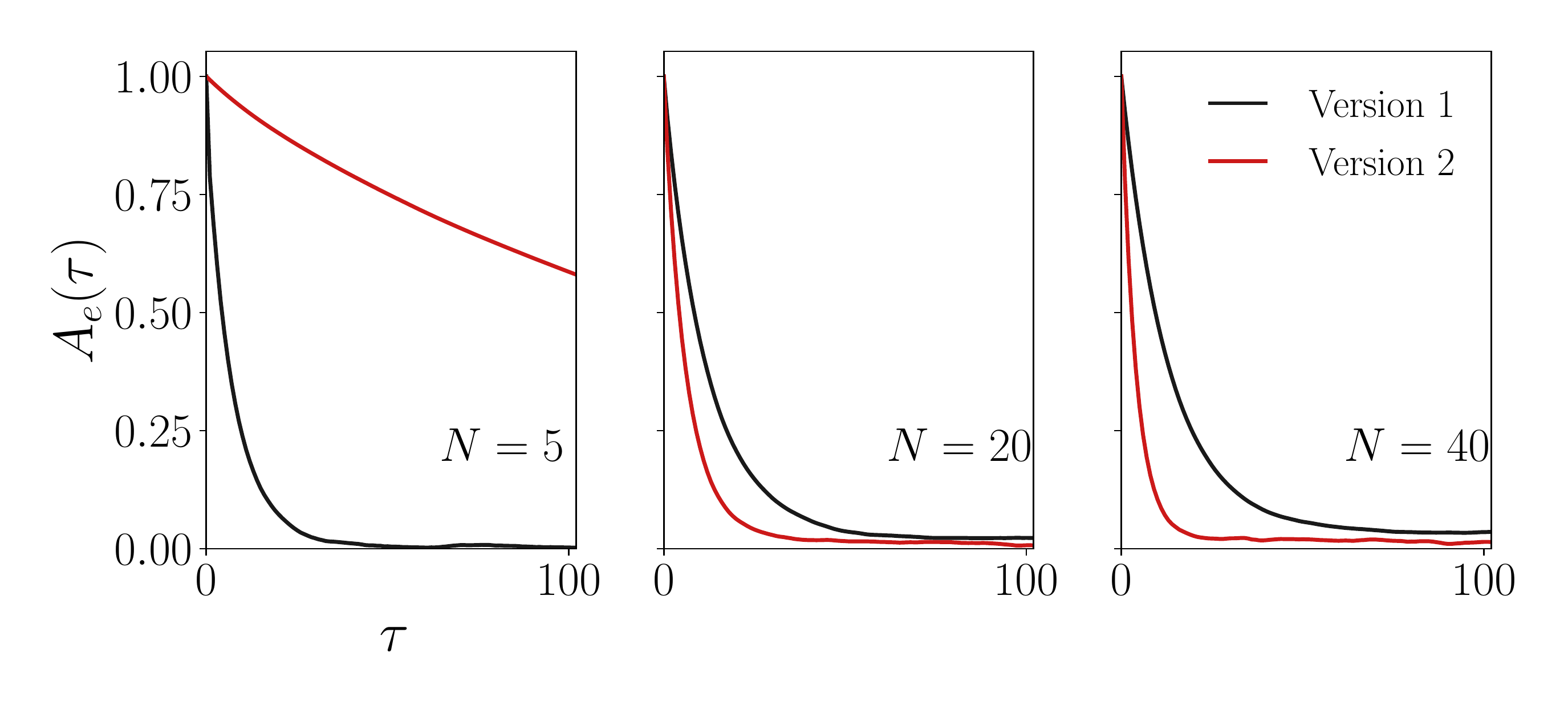}
 \caption{\label{enrautocorr_comparison_v1v2_L16} Autocorrelation of the energy per unit site on a $16 \times 16$ lattice measured after each Monte Carlo step for the two versions of the algorithm for $N=5,\, 20,\, 40$. Version 1 does better for relatively smaller $N$.}
\end{figure}

\section{Observables}

\begin{figure}
 \subfigure[]{\includegraphics[width=0.49\linewidth]{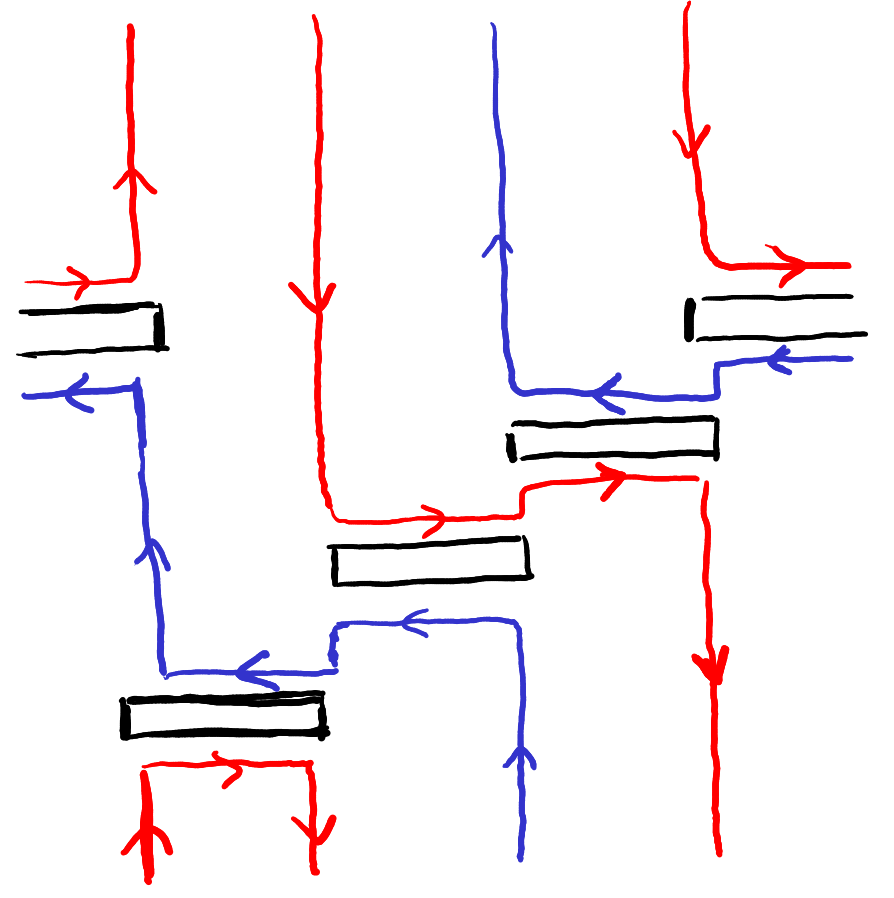}}
 \subfigure[]{\includegraphics[width=0.49\linewidth]{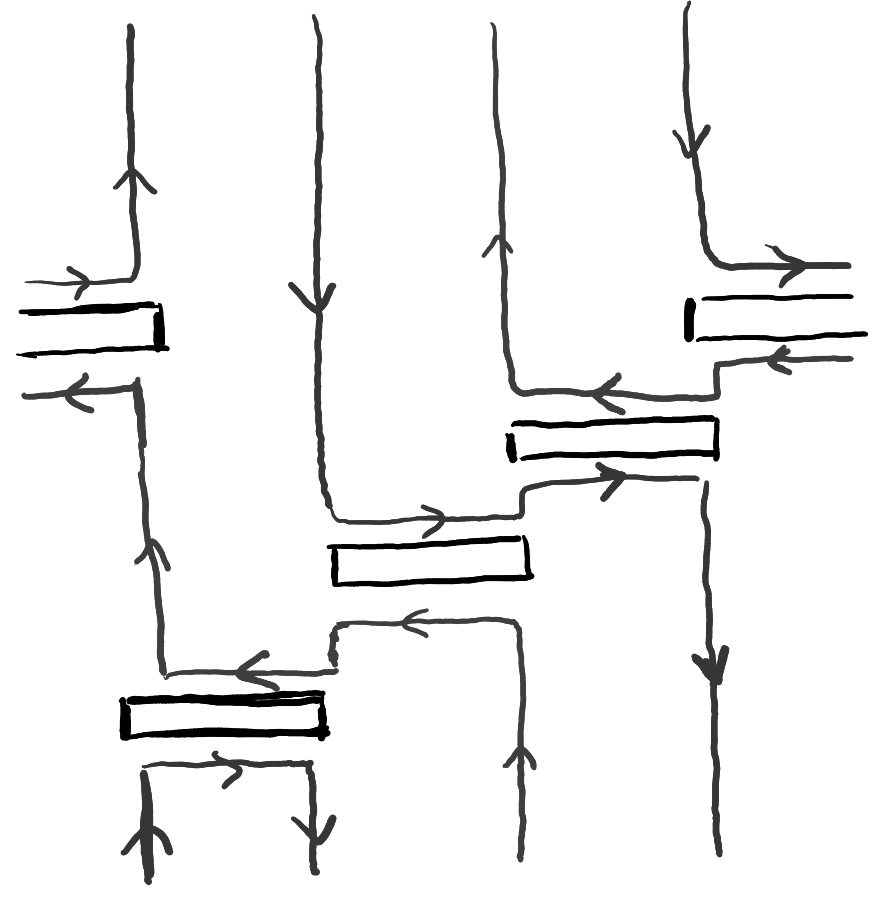}}
 \caption{\label{fig:winding_conf}A configuration of a pair of winding loops in (a) SSE and (b) resummed-SSE. In SSE, contributions to winding come from a string of off-diagonal operators. In such configurations, loops of two different colors touch the operators on the two sides of each operator vertex along the string of operators. Hence, winding loops always appear in pairs, each winding in opposite directions.}
\end{figure}

\emph{Spin stiffness:} As mentioned in the main manuscript, the computation of spin stiffness in this algorithm is different from the computation in standard SSE. In standard SSE, the stiffness is computed using the winding of colored loops~\cite{Sandvik_2010_review,sandvik1997:prb}. For every off-diagonal operator, two differently colored loops touch both sides of the operator vertex. Each of these loops contribute to winding of their respective colors with opposite sign at this vertex. If these loops instead have the same color, i.e. the operator is diagonal, the contribution to winding of that color cancels out. Winding loops occur in pairs with opposite signs whenever there is a string of off-diagonal operators as shown in Fig.~\ref{fig:winding_conf}(a).

In our algorithm, there is no notion of color or diagonal/off-diagonal bonds. Therefore we proceed as 
follows: we seek to find the fraction of colored loops, upon coloring back the
uncolored loop configuration, that would have contribute to stiffness or (colored) winding fluctuations as described in the previous paragraph. 
Consider a pair of winding loops, $l_1$ and $l_2$, with a signed winding of 
$\mathcal{W}$ and $-\mathcal{W}$ respectively (the
number of times they wind around the spatial periodic boundary condition;
assigning a sign to the winding is possible on bipartite lattices since
the loops are ``orientable" as shown in Fig.~\ref{fig:winding_conf})
in some configuration. 
Now, imagine coloring this pair of winding loops.
Thus, the winding of color $\alpha$ in terms of the uncolored loop windings ($\mathcal{W}$) for this pair of loops $l_1$ and $l_2$  is:
\begin{equation}
 \mathcal{W}_{\alpha} = [n_{\alpha}(l_1)-n_{\alpha}(l_2)]\mathcal{W} 
\end{equation}
where $n_{\alpha}(l)=1$ if loop $l$ is of color $\alpha$, otherwise  $n_{\alpha}(l)=0$. It is clear that $\mathcal{W}_{\alpha}$ is non-zero only when exactly one of the loops has color $\alpha$. The number of ways in which exactly one of the two loops has color $\alpha$ is $2(N-1)$. If one now averages over all possible colorings of $l_1$ and $l_2$, the fraction of loop colorings that contribute to winding of color $\alpha$ is $\frac{2(N-1)}{N^2}$. Therefore:
\begin{equation}
 \overline{ (\mathcal{W}_{\alpha})^2}  = \frac{2(N-1)}{N^2}
 \mathcal{W}^2
 \label{eq:walpha2}
\end{equation}
where $\overline{ \mathcal{W}_{\alpha}^2}$ denotes the average of $\mathcal{W}_{\alpha}^2$ over all possible loop colorings of $l_1$ and $l_2$. We further average over the colors (for better statistics) to arrive at $\overline{ \mathcal{W}^2_c } = \frac{1}{N} \sum_{\alpha} \overline{ \mathcal{W}_{\alpha}^2 }$. The uncolored winding squared for $l_1$ and $l_2$ is simply $\mathcal{W}^2_u=2 \times \mathcal{W}^2$. Note that the factor in Eq.~\ref{eq:walpha2} are just a function of $N$ and thus remains the same for all pairs of winding loops. Therefore, the squared winding of uncolored loops in terms of the average squared winding of colored loops  is given by the following relation:
\begin{equation}
 \mathcal{W}_u^2 = \frac{N^2}{(N-1)} \overline {\mathcal{W}_c^2}   
\end{equation}

\emph{VBS order parameter}:  In the VBS phase, the Fourier transform of the equal time correlator of the bond operator $B_\lambda(\mathbf{r})$ as defined in the main text, $\tilde{C}_{\phi^2_\lambda}(\mathbf{k})=\frac{1}{N_s} \sum_{\mathbf{r}} e^{i \mathbf{k}.\mathbf{r}} \langle B_\lambda({\mathbf{0}})B_\lambda({\mathbf{r}})\rangle$ has a Bragg peak at $\mathbf{k}=\pi \hat{e}_\lambda$, where $\hat{e}_x=(1,0)$ and $\hat{e}_y=(0,1)$. The height of this peak is the square of the VBS order parameter, $\langle \phi^2_\lambda \rangle =\tilde{C}_{\phi^2_\lambda}(\pi \hat{e}_\lambda)$. The correlator of the bond operator as defined in the main manuscript is measured in the same way as in standard SSE.

\begin{figure}[t]
 \includegraphics[width=\linewidth,clip=true,trim=0 12mm 0mm 0mm]{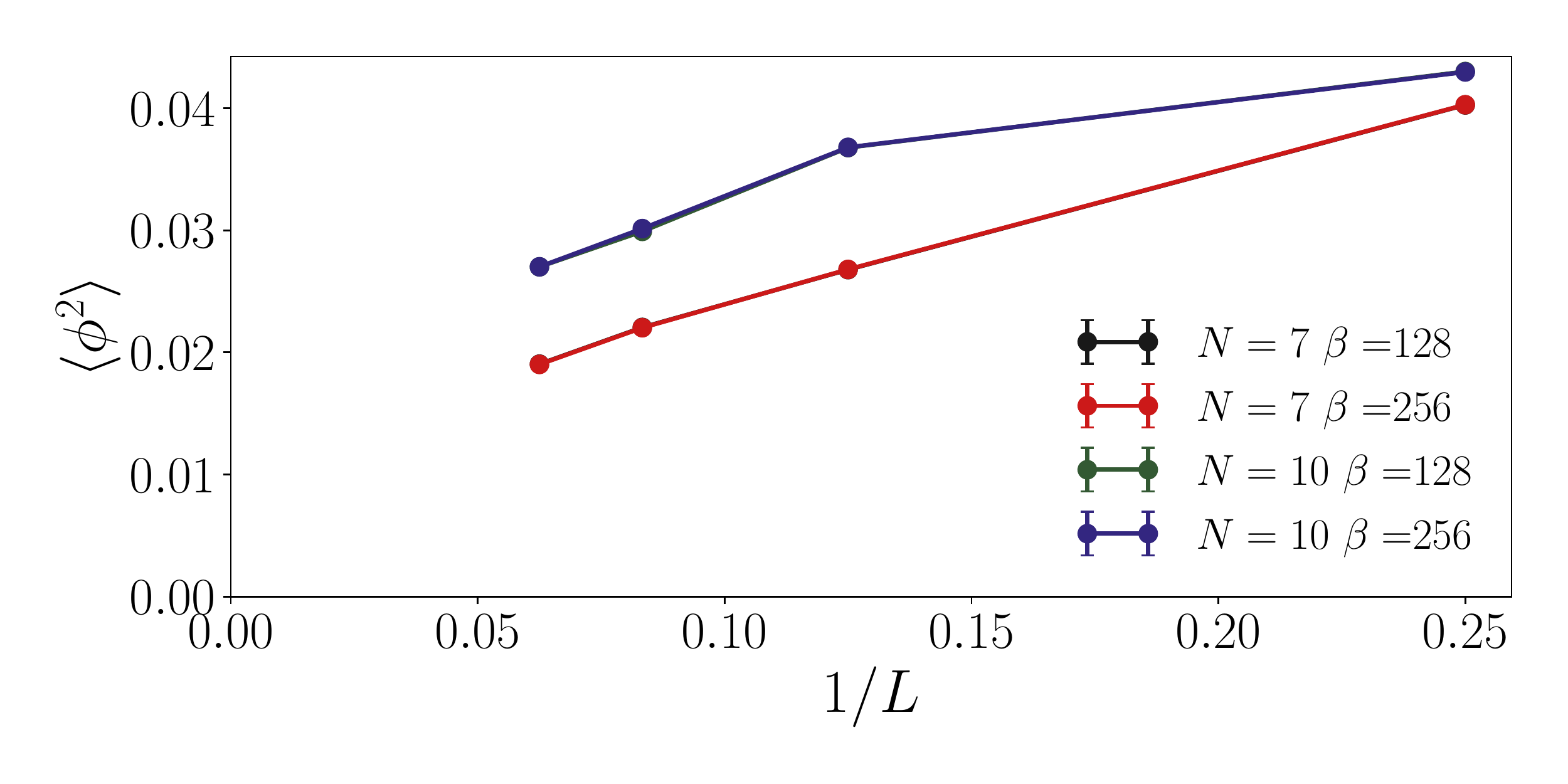}
 \caption{\label{suppfig:betaconvergence} The VBS order parameter, $\langle \phi^2 \rangle$, plotted as a function of inverse system size for two different inverse temperatures, $\beta=128$ and $\beta=256$. The curves for $\beta=128$ and $\beta=256$ lie right on top of each other showing convergence with temperature. }
\end{figure}

\section{Temperature dependence of Observables}
As highlighted in the main text, 
our algorithm is a 
finite-$T$ generalization of valence bond $T=0$ projector QMC algorithm~\cite{Sandvik_2005,Sandvik_2010b}
and its continuous-$N$ variant~\cite{Beach_2009}
but with periodicity in imaginary time 
and without being in total $S=0$ sector. 
Our algorithm is a generalization in the sense that it retains 
the (uncolored) ``valence bond'' spirit even though not being 
restricted to the total $S=0$ sector.  
There can for example be uncolored loops that 
loop non-trivially along the imaginary time direction (e.g. on sites 1 and 4 in
Fig.~\ref{suppfig:diag1}.
At low enough temperatures in an antiferromagnet, such time-winding loops
however will be extremely unlikely even though in principle allowed,
which will practically keep the ensemble in total $S^z=0$ sector.
In the next section, we will see that these cases actually correspond to ``monomer'' defects in valence bond configurations.

We now show how finite temperature data converges towards zero temperature.
We choose $\beta=128\, J$ for studying groundstate properties at $N = 7, 10, 13$. Fig.~\ref{suppfig:betaconvergence} shows convergence of the VBS order parameter as a function of temperature for $N = 7, 10$.

\section{Approach to QDM  Hilbert space at large $N$}
\begin{figure}
\includegraphics[width=0.8\linewidth]{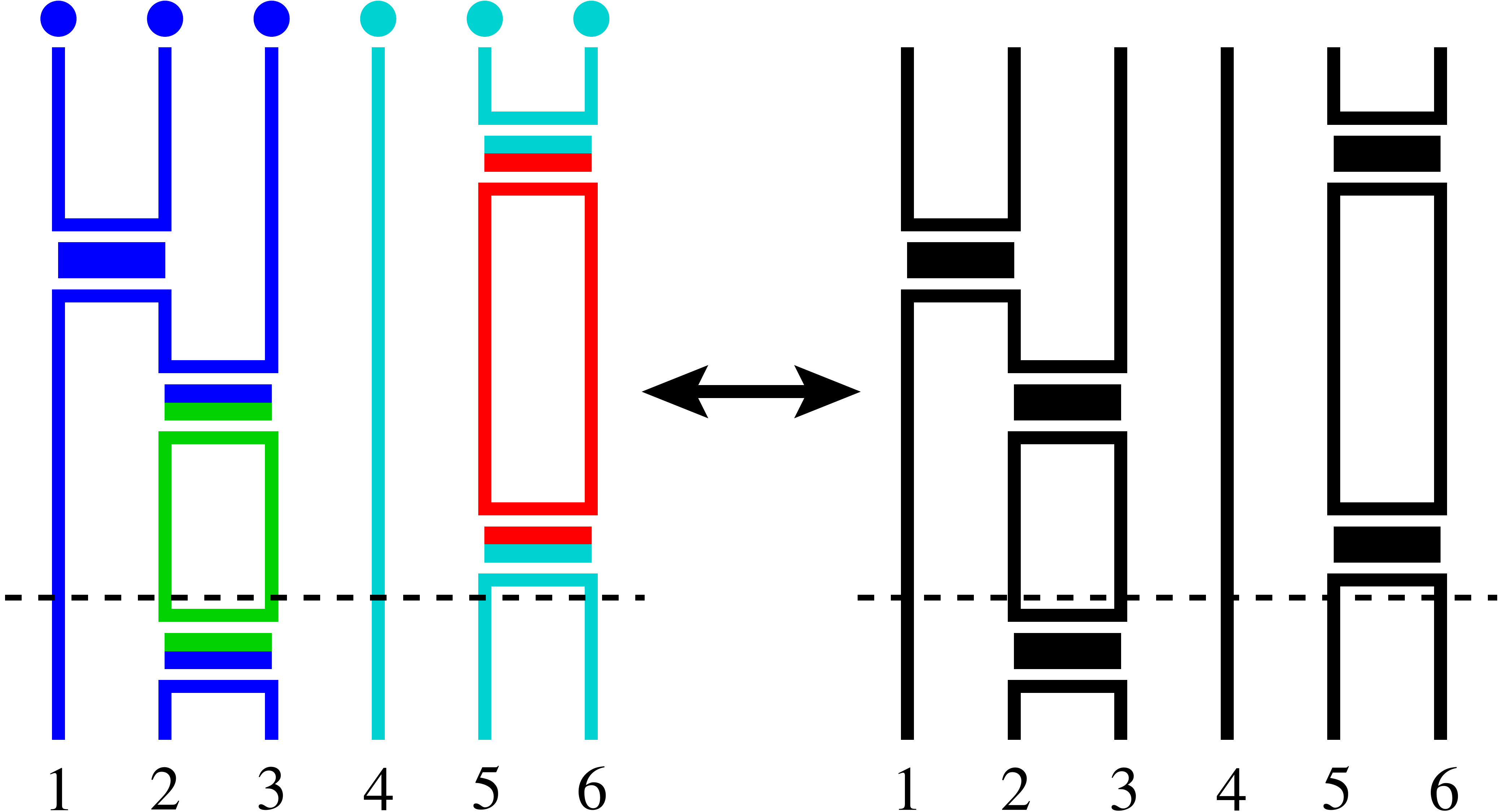}
\caption{\label{suppfig:diag1} An illustrative operator string configuration
for a) standard SSE, and b) for resummed SSE. The dashed horizontal line
depicts a typical time slice in the imaginary time direction.
}
\end{figure}

For this discussion, we start by asking what each time slice represents
in the uncolored loop ensemble. In the colored loop ensemble, each time
slice represents a definite spin state, i.e. each site of the lattice
has a definite spin or color  $\prod_i \otimes |\alpha_i\rangle$ at all time slices.
After resumming, the uncolored loop is a ``superposition" over the colors
in space-time. What is the effect of this superposition at
some chosen time slice. Each time slice makes a slice perpendicular to the
imaginary time direction in the uncolored loop gas configuration
as shown by the dashed line in Fig.~\ref{suppfig:diag1}. Under this,
the lattice gets subdivided into a disjoint sets of sites with the following
properties: a) two sites from
the same set are connected by the same uncolored loop, and
b) two sites from two different
sets are not connected by any loop. 
Thus, sites from the same set upon coloring back will necessarily have the same color,
while sites belonging to different sets need not have the same color,
as can be seen on the left panel of Fig.~\ref{suppfig:diag1}.
There can be sets with only one site at finite-$T$, e.g. sites 1 and 4 in
Fig.~\ref{suppfig:diag1} at the dashed time slice.

Any given set of sites with the same color upon coloring back
may now be interpreted as follows: It is the \emph{overlap} of two
appropriate states. This is essentially a generalization of the 
``central" time slice interpretation in the
$T=0$ valence bond projector QMC as an overlap of two valence bond states
while stochastically estimating various physical observables.
(e.g. see the central time slice represented as a dashed line 
in Fig. 91 of Ref.~\cite{Sandvik_2010_review}).
For the base cases of 
\begin{itemize}
\item 1) a set with one site ($i$),
the two appropriate states could be 
$|s^x_i=\frac{1}{2}\rangle \equiv \frac{1}{\sqrt{2}}
 \left( |s^z_i=\frac{1}{2}\rangle + |s^x_i=-\frac{1}{2}\rangle\right)$ or
any other state in the $xy$-plane of the Bloch sphere, suitably
generalized to any $N$ (say $|\bullet_i\rangle \equiv \frac{1}{\sqrt{N}} 
\sum_\alpha |\alpha_i\rangle$).
Upon overlapping them $\langle \bullet_{i}|\bullet_{j}\rangle$,
we indeed get the same color upon coloring back at this single site 
with each coloring
being equally weighted, i.e. $\langle \bullet_{i}|\bullet_{j}\rangle = 
\frac{1}{N} \sum_\alpha \langle \alpha_i|\alpha_i\rangle$
This case arises because the uncolored loop ensemble is
at finite-$T$. 

\item 2) a set with two sites ($i$ and $j$, not necessarily nearest
neighbor), the two appropriate states are singlets on the bond formed
by $i$ and $j$, $|$\textbf{---}$_{ij}\rangle \equiv \frac{1}{\sqrt{N}} 
\sum_\alpha |\alpha_i \alpha_j \rangle$.
Upon overlapping them $\langle $\textbf{---}$_{ij}|$\textbf{---}$_{ij}\rangle$,
we indeed get the same color upon coloring back at this set of two sites with each coloring
being equally weighted, i.e. 
$\langle $\textbf{---}$_{ij}|$\textbf{---}$_{ij}\rangle =
\frac{1}{N} \sum_\alpha \langle \alpha_i \alpha_j | \alpha_i \alpha_j\rangle$.

\end{itemize}

\begin{figure}[t]
\includegraphics[width=0.8\linewidth]{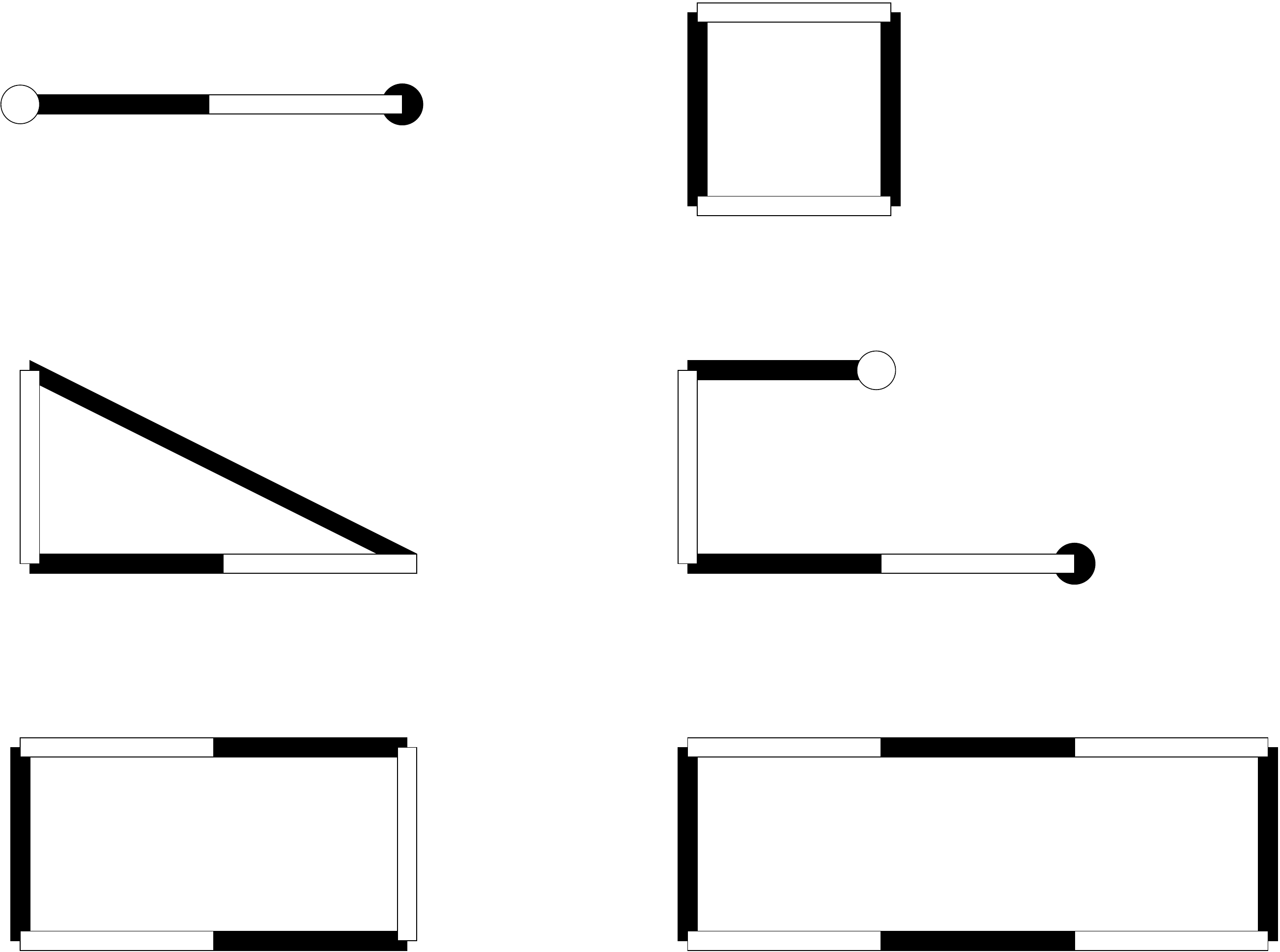}
\caption{ \label{suppfig:diag2}
Examples of disjoint sets with more than two sites.
The top left and middle row are examples that violate the QDM Hilbert
space condition, while the other examples respect this condition.
}
\end{figure}

\begin{figure*}[t]
\includegraphics[width=0.45\linewidth]{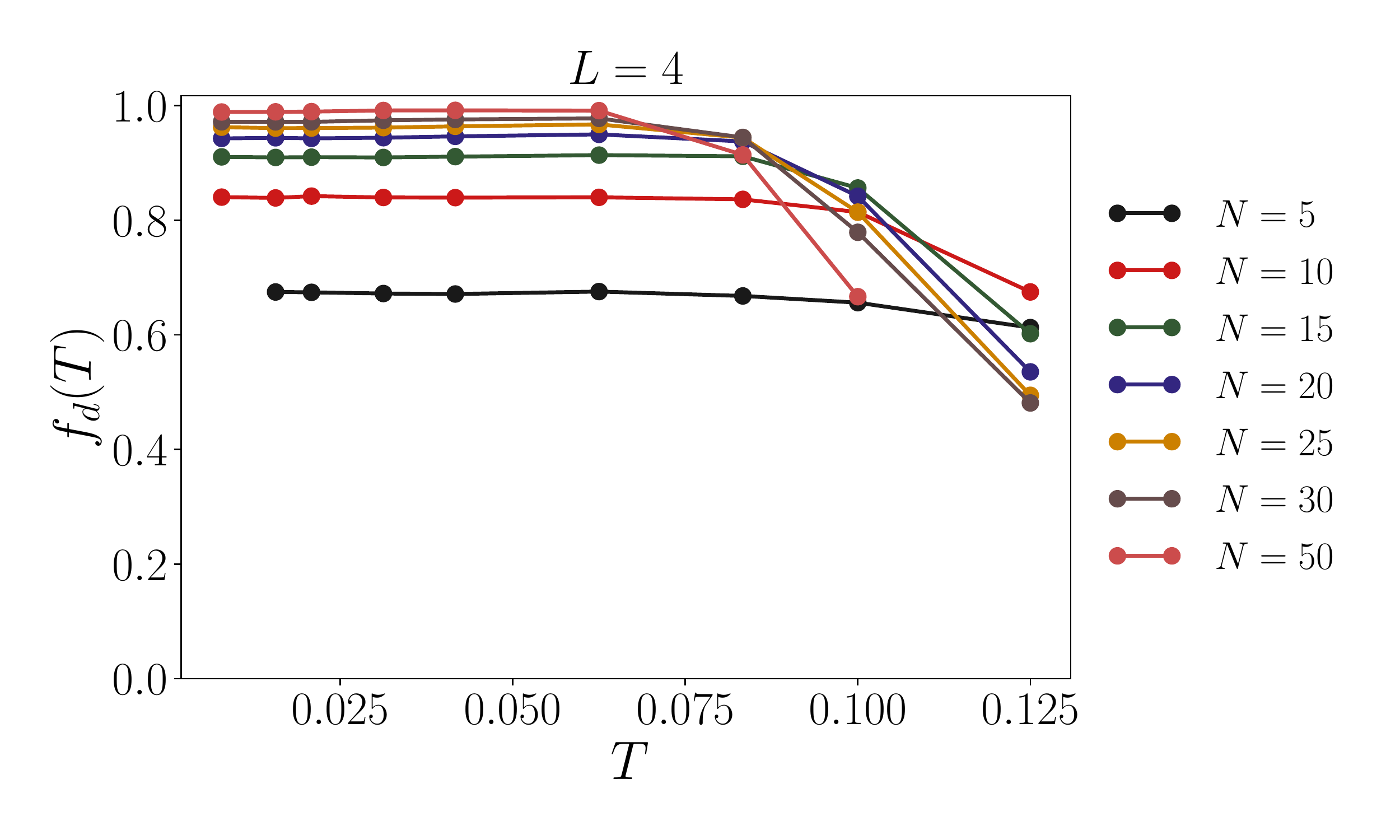}
\includegraphics[width=0.45\linewidth]{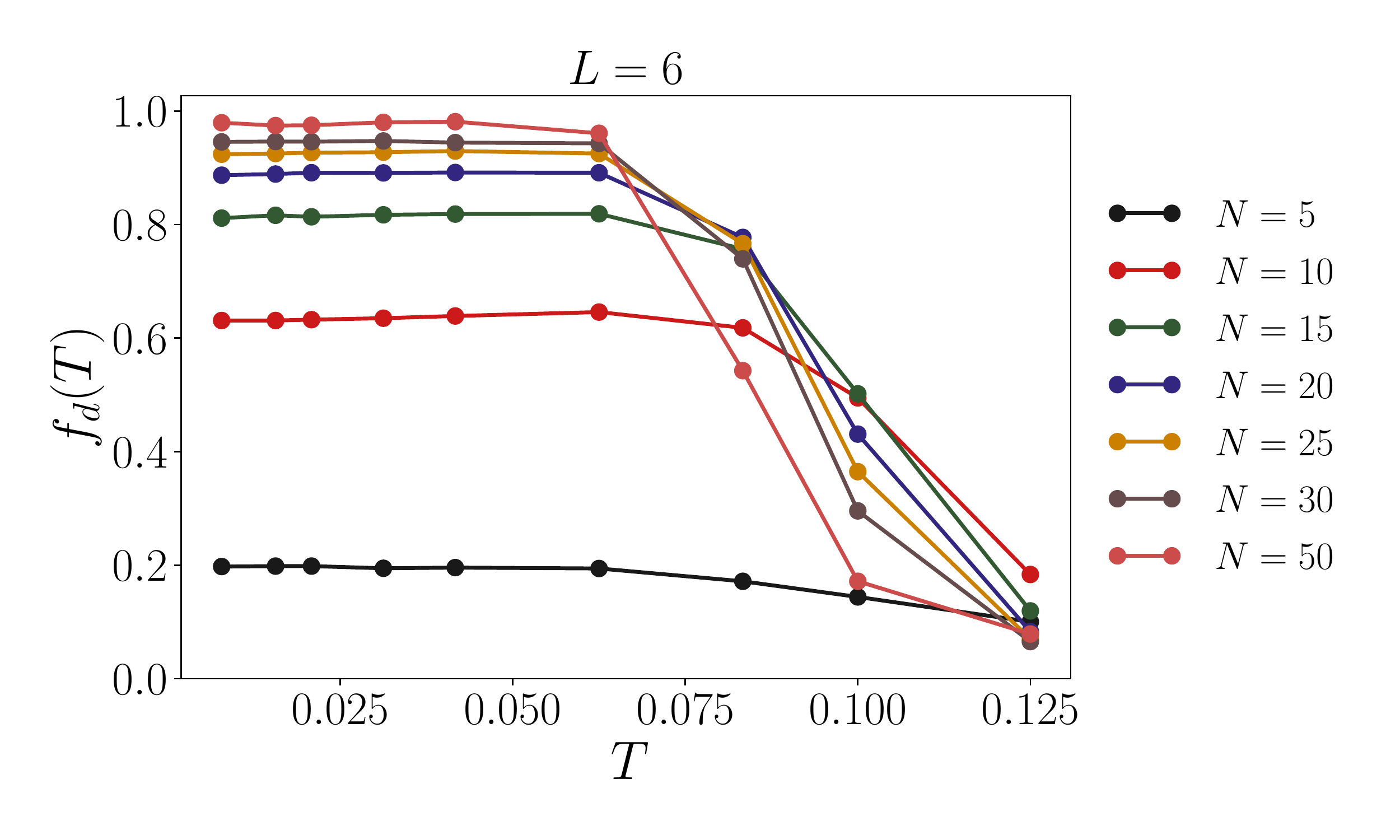}
\includegraphics[width=0.45\linewidth]{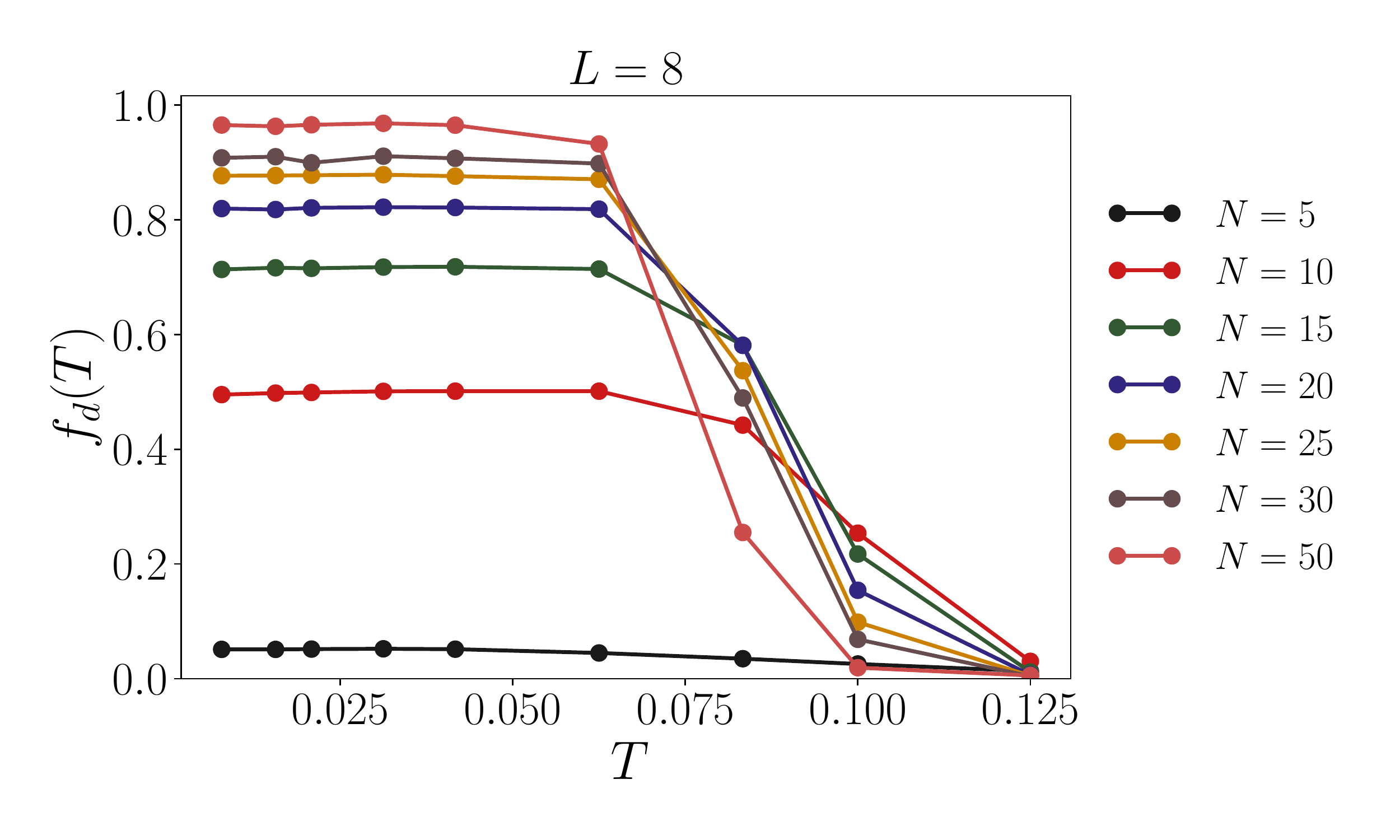}
\caption{ \label{suppfig:qdm_stats}
The fraction $f_d$ of typical time slices which are consistent with the
QDM Hilbert space is shown above vs. temperature $T$ for several values
of $N$. We see that there is a non-negligible violations of the QDM
Hilbert space condition already at $L=8$ for moderately large values
of $N$. This suggests that approaching the perturbative neighborhood
of the QDM Hilbert space at large system sizes for $SU(N)$ magnets 
might require really large values of $N$. Formally, this approach is
expected for $N \gg 1$.
}
\end{figure*}
This interpretation carries over to sets with more number of sites.
They correspond to a ``transposition graph" due to the overlap
of two dimer or singlet configurations that are compatible with
the geometry of the set of sites.
From this point of view, the first point in the list above on the case
of a set with a single site corresponds to a monomer defect at the
corresponding time slice.
Several examples of such disjoint sets with more than two sites
interpreted as overlaps of two singlet configurations 
are shown in Fig.~\ref{suppfig:diag2}. 
For the case of 
the top left example in Fig.~\ref{suppfig:diag2}, the two appropriate states
are the following configurations with monomers, 
$|$\textbf{---}$_{ij}\rangle \otimes |\bullet_{k}\rangle$
and 
$|\bullet_{i}\rangle \otimes |$\textbf{---}$_{jk}\rangle$.
$i,j,k$ is, say, a linear indexing from left to right 
of the three sites in this example,
and the black monomer and dimer represents one of these two states, while the white
monomer and dimer represents the other state. 
Again, upon overlapping these two states,
we indeed get the same color upon coloring back at this set of three sites with each coloring
being equally weighted, i.e.
$\big( \langle $\textbf{---}$_{ij}\rangle \otimes \langle \bullet_{k} | \big)
| \big( |\bullet_{i}\rangle \otimes |$\textbf{---}$_{jk}\rangle \big)
=
\frac{1}{N^2} \sum_\alpha \langle \alpha_i \alpha_j \alpha_k 
| \alpha_i \alpha_j \alpha_k \rangle$.
For the case of 
top right example in Fig.~\ref{suppfig:diag2}, the two appropriate states
are the following singlet configurations, 
$|$\textbf{---}$_{ij}\rangle \otimes |$\textbf{---}$_{kl}\rangle$
and 
$|$\textbf{---}$_{kl}\rangle \otimes |$\textbf{---}$_{li}\rangle$.
$i,j,k,l$ is, say, a clockwise indexing of the four sites in this example,
and the black dimers represent one of these two states, while the white
dimers represent the other state. 
Again, upon overlapping these two states,
we indeed get the same color upon coloring back at this set of four sites with each coloring
being equally weighted, i.e.
$\big( \langle $\textbf{---}$_{ij}\rangle \otimes \langle $\textbf{---}$_{kl} | \big)
| \big( $\textbf{---}$_{jk}\rangle \otimes |$\textbf{---}$_{li}\rangle \big)
=
\frac{1}{N^2} \sum_\alpha \langle \alpha_i \alpha_j \alpha_k \alpha_l
| \alpha_i \alpha_j \alpha_k \alpha_l \rangle$. For sets with even more sites
as in the other examples of Fig.~\ref{suppfig:diag2}, a similar interpretation
can be applied by generalizing the above steps appropriately.

As mentioned earlier, this is quite analogous to the time slice interpretation in the
$T=0$ valence bond projector QMC. In the standard implementation
of $T=0$ valence bond projector QMC where one does not have any periodicity
in imaginary time but rather ``open" time boundary conditions, 
one usually does the updates in the spin basis
using standard SSE updates \emph{a la} Sandvik and Evertz. 
We are extending the same time slice interpretation to the finite-$T$
case now with periodicity in imaginary time such that all time
slices have the same status. Then,
the corollary follows that
for the case of $T=0$ as well, 
one can directly work with the uncolored loops, if so one chooses,
using the resummation-based updates that is the subject of this paper.

Based on the above interpretation, we now quantify the approach
to the QDM Hilbert space as follows: after 
equilibration or warm-up phase of the QMC simulation, 
for a typical time slice of the uncolored loop configuration,
we construct the disjoint sets with sites that lie on the same uncolored loop.
Then, we check if each disjoint set is compatible with the overlap of
two nearest-neighbor dimer tilings or not, since the standard 
QDM Hilbert space contains only nearest-neighbor dimers with the
constraint that no two dimers can meet at a site. If all the disjoint
sets at a typical time slice are compatible with such nearest-neighbor
dimer tiling overlaps, then the time slice is in the QDM Hilbert space.
We collect statistics on this for several sizes and values of $N$ as
shown in Fig.~\ref{suppfig:qdm_stats}. We see that uncolored loop ensemble
starts approaching the QDM Hilbert space as $N$ gets large, and $T$ gets
small. However, we note that there is a system size dependence to this,
and for thermodynamically large systems, one might have to go to really
large $N$ or really low $T$ to be in the perturbative neighborhood of the QDM
Hilbert space. 
E.g., for $N=15$ and $L=8$, there is already a 30\% violation to the
QDM Hilbert space condition, and the severity of such violations will
only get worse as system size increases simply due to entropic
reasons, i.e. there are more regions available in the lattice as system
size increases where violations may occur.

\bibliography{resum_supp_refs.bib,resum_refs.bib}